\documentclass[manuscript]{acmart}

\AtBeginDocument{%
  }

\copyrightyear{2022}
\acmYear{2022}
\setcopyright{acmcopyright}\acmConference[ICMI '22]{INTERNATIONAL CONFERENCE
ON MULTIMODAL INTERACTION}{November 7--11, 2022}{Bengaluru, India}
\acmBooktitle{INTERNATIONAL CONFERENCE ON MULTIMODAL INTERACTION (ICMI
'22), November 7--11, 2022, Bengaluru, India}
\acmPrice{15.00}
\acmDOI{10.1145/3536221.3556630}
\acmISBN{978-1-4503-9390-4/22/11}
\usepackage{algorithm}
\usepackage{algorithmic}

\acmSubmissionID{123-A56-BU3}

\begin{document}

\title{Make Acoustic and Visual Cues Matter: CH-SIMS v2.0 Dataset and AV-Mixup Consistent Module}
\author{Yihe Liu$^{1 2 3}$*,\quad Ziqi Yuan$^{1 3}$*,\quad Huisheng Mao$^{1 3}$, \quad Zhiyun Liang$^{1 5}$, \quad Wanqiuyue Yang$^{1 4}$, \quad Yuanzhe Qiu$^{1 2 3}$, \quad Tie Cheng$^{1 4}$, \quad Xiaoteng Li$^{1 2 3}$, \quad Hua Xu$^{1\dagger}$,\quad Kai Gao$^{2\dagger}$\\}

% \author{YH Liu$^{1 2 3}$*,\quad ZQ Yuan$^{1 2}$*,\quad HS Mao$^{1 2}$, \quad ZY Liang$^{5}$, \quad WQY Yang$^{4}$, \quad YZ Qiu$^{1 2 3}$, \quad T Cheng$^{4}$, \quad XT Li$^{1 2 3}$, \quad H Xu$^{1\dagger}$,\quad K Gao$^{2\dagger}$}

\makeatletter
\def\authornotetext#1{
\if@ACM@anonymous\else
    \g@addto@macro\@authornotes{
    \stepcounter{footnote}\footnotetext{#1}}
\fi}
\makeatother
\authornotetext{Equal Contribution.}
\authornotetext{Corresponding author.}

\affiliation{
 \institution{\textsuperscript{\rm 1}State Key Laboratory of Intelligent Technology and Systems, Department of Computer Science and Technology, Tsinghua University}
 \city{\textsuperscript{\rm }Beijing}
 \country{\textsuperscript{\rm }China}
 \postcode{\textsuperscript{\rm }100084}
 }
 \affiliation{
 \institution{\textsuperscript{\rm 2}School of Information Science and Engineering, Hebei University of Science and Technology}
 \city{\textsuperscript{\rm }Shijiazhuang}
 \country{\textsuperscript{\rm }China}
 \postcode{\textsuperscript{\rm }050018}
 }
 \affiliation{
 \institution{\textsuperscript{\rm 3}Beijing National Research Center for Information Science and Technology(BNRist)}
 \city{\textsuperscript{\rm }Shijiazhuang}
 \country{\textsuperscript{\rm }China}
 \postcode{\textsuperscript{\rm }050018}
 }
 \affiliation{
 \institution{\textsuperscript{\rm 4}Beijing University of Posts and Telecommunications}
 \city{\textsuperscript{\rm }Beijing}
 \country{\textsuperscript{\rm }China}
 \postcode{\textsuperscript{\rm }100876}
 }
 \affiliation{
 \institution{\textsuperscript{\rm 5}China Agricultural University}
 \city{\textsuperscript{\rm }Beijing}
 \country{\textsuperscript{\rm }China}
 \postcode{\textsuperscript{\rm }100091}
 }
 
\email{512796310@qq.com}
\email{yzq21@mails.tsinghua.edu.cn}
\email{xuhua@tsinghua.edu.cn}

\begin{abstract}
\section*{Abstract}
  Multimodal sentiment analysis (MSA), which supposes to improve text-based sentiment analysis with associated acoustic and visual modalities, is an emerging research area due to its potential applications in Human-Computer Interaction (HCI). However, existing researches observe that the acoustic and visual modalities contribute much less than the textual modality, termed as text-predominant. Under such circumstances, in this work, we emphasize making non-verbal cues matter for the MSA task. Firstly, from the resource perspective, we present the CH-SIMS v2.0 dataset, an extension and enhancement of the CH-SIMS. Compared with the original dataset, the CH-SIMS v2.0 doubles its size with another 2121 refined video segments containing both unimodal and multimodal annotations and collects 10161 unlabelled raw video segments with rich acoustic and visual emotion-bearing context to highlight non-verbal cues for sentiment prediction. Secondly, from the model perspective, benefiting from the unimodal annotations and the unsupervised data in the CH-SIMS v2.0, the Acoustic Visual Mixup Consistent (AV-MC) framework is proposed. The designed modality mixup module can be regarded as an augmentation, which mixes the acoustic and visual modalities from different videos. Through drawing unobserved multimodal context along with the text, the model can learn to be aware of different non-verbal contexts for sentiment prediction. Our evaluations demonstrate that both CH-SIMS v2.0 and AV-MC framework enable further research for discovering emotion-bearing acoustic and visual cues and pave the path to interpretable end-to-end HCI applications for real-world scenarios. The full dataset and code are available for use at \url{https://github.com/thuiar/ch-sims-v2}.
\end{abstract}

\begin{CCSXML}
<ccs2012>
<concept>
<concept_id>10002951.10002952.10002953.10010820.10002958</concept_id>
<concept_desc>Information systems~Semi-structured data</concept_desc>
<concept_significance>500</concept_significance>
</concept>
<concept>
<concept_id>10002951.10003227.10003251</concept_id>
<concept_desc>Information systems~Multimedia information systems</concept_desc>
<concept_significance>300</concept_significance>
</concept>
<concept>
<concept_id>10002951.10003317.10003347.10003353</concept_id>
<concept_desc>Information systems~Sentiment analysis</concept_desc>
<concept_significance>500</concept_significance>
</concept>
</ccs2012>

\end{CCSXML}



\maketitle

\section{Introduction}

Understanding the speakers' sentiment is a crucial step for intelligent Embodied Conversational Agents (ECAs) in generating the empathetic responses \cite{ghosal2020utterance, ma2020survey}. Traditional embodied agents analyze the users' sentiment with unimodal textual information, inevitably confronting performance gaps due to text ambiguity and irony \cite{poria2020beneath}. With advanced micro-sensors, multiple sensory resources, such as visual and vocal, can be recorded along with the spoken words and lead to the Multimodal Sentiment Analysis (MSA) task \cite{tao2005affective, baltruvsaitis2018multimodal, poria2018multimodal, liang2021multibench}, which aims to judge the users' sentiment using textual, acoustic, and visual behaviors. 

Until now, however, previous MSA benchmarks have still over-reliance on the textual modality. In literature \cite{li2020multimodal, misa, luo2021scalevlad}, researchers report about a 30\% binary accuracy drop when removing the textual modality on MOSI dataset (80\%+ with text, while about 54\% without text). In this work, we term the above phenomenon as \textit{text-predominant}. Such an underestimate of acoustic and visual behaviors violates the motivation of integrating multimodal resources and seriously limits its applications in real-world scenarios where textual modality is always imperfect due to the potential ASR error \cite{wu2022sentiment, amiriparian2021impact}. As a result, effectively exploiting non-verbal cues becomes the most crucial challenge in the field of multimodal sentiment analysis. The reasons for the text-predominant can be summarized from two aspects:

\noindent\textbf{Resource Limitations.} On the one hand, resource size and quality are essential but commonly ignored factors. Firstly, for raw video collection, the major concern in MSA resources lies in the scarcity of emotion-bearing acoustic and visual behaviors. Although some of the existing MSA corpus, such as MOSEI \cite{mosei}, have reached the level of 20,000 in scale, its emotion-bearing acoustic and visual behaviors are still limited due to the ubiquitous video's blurry and flat tone. Moreover, the obvious label bias in previous MSA corpus (MOSEI \cite{mosei} contains 69\% positive instances and CH-SIMS contains 69\% negative instances) results in trivial unimodal solutions, which further prevents acoustic and visual representation learning. Secondly, for data annotation, as stated in literature \cite{yu2020ch}, the unified multimodal annotations can be inconsistent with the independent sentiment of single modality. Thus, unimodal annotation is necessary for the evaluation of unimodal representation learning and is assumed to benefit the fine-grained sentiment intensity prediction. In general, the a high-quality MSA datasets should have a certain scale, consist of diverse nonverbal context information, balance instance annotation distribution, and contain unimodal annotation.

\noindent\textbf{Under-optimized Acoustic and Visual Representation.}
In addition to the resource limitations, the under-optimized acoustic and visual representation in a joint learning framework is another factor for the text-predominant phenomenon \cite{peng2022balanced}. Previous MSA benchmarks focus more on the study of the multimodal fusion process and commonly utilize a joint representation learning framework to learn a unified multimodal representation \cite{ tfn, mfn, mult}. As stated in literature \cite{misa}, benefited from pretrained language model, directly utilizing the emotion cues from text modality is much easier than exploring the sparse emotion-bearing non-verbal cues. The few contributions of the acoustic and visual modality to the joint fusion result might prevent the training of the acoustic and visual encoder. Although previous work report their multimodal result outperforms each unimodal result, they still fail to fully utilise the acoustic and visual modalities. As a result, there is an emerging trend to manually introduce other supervision or perform gradient modulation for acoustic and visual modality to improve the non-verbal representation learning \cite{yu2020ch, self-mm, peng2022balanced}.

In this work, for the challenge of MSA resources, we construct the CH-SIMS v2.0 dataset, an enhancement and extension of the CH-SIMS, which includes 4402 supervised data with unimodal annotation and over 10,000 unsupervised data. The CH-SIMS v2.0 dataset is collected from 11 different scenarios to simulate real-world HCI scenarios paving the path for ECAs to understand users' emotions. Some of the typical instances are shown in Figure \ref{fig: CH-SIMS}. For the challenge of acoustic and visual representation learning, we introduce the Mixup \cite{zhang2017mixup} strategy into the MSA task taking advantage of the provided unimodal annotations to aware the model of the various non-verbal behaviors. The main contributions of this work are summarized below.
\begin{itemize}
\item[1)] From a resource perspective, CH-SIMS v2.0, the largest semi-supervised Chinese MSA dataset with diverse non-verbal behaviors is built for HCI community to explore the effectiveness of acoustic and visual behaviors.

\item[2)] From a representation learning perspective, the Acoustic Visual Mixup Consistent (\textbf{AV-MC}) framework is designed to simulate the potential unobserved non-verbal behaviors corresponding to the same spoken words through a mixup strategy. Moreover, the proposed AV-MC framework can be easily adapted to both supervised and semi-supervised training paradigms.

\item[3)] Empirically, this paper conducts extensive experiments on CH-SIMS v2.0 which serve as the baselines and intuitive demonstration showing the significance of making full use of non-verbal behaviors.

\end{itemize}

\section{Related Works}

% Multimodal sentiment analysis attracts more and more attention in recent years with the popularity of the microsensors equipped with ECAs. In this section, we first introduce the existing research in the MSA area, including datasets in Section \ref{sec: rel_data}, and approaches in Section \ref{sec: rel_model}. Finally, we summarized recent efforts in the Mixup strategy in Section \ref{sec: rel_mix}.

\subsection{Multimodal Sentiment Analysis Dataset}
\label{sec: rel_data}
As the demand for emotion aware applications increases, various multimodal sentiment datasets have been constructed in the past few years including \textbf{IEMOCAP} \cite{busso2008iemocap}, \textbf{YouTuBe} \cite{morency2011towards}, \textbf{MOSI} \cite{mosi}, \textbf{MOSEI} \cite{mosei}, \textbf{MELD} \cite{2019MELD}, and the latest \textbf{MOSEAS} \cite{zadeh2020cmu}. Most of the above datasets contain only English resources, however, the expression of emotion in different cultures and native languages can be various. For Chinese MSA task, \citet{li2017cheavd} propose the first Chinese emotional acoustic-visual dataset, \textbf{CHEAVD}. Recently, \citet{yu2020ch} propose the \textbf{CH-SIMS}, a Chinese MSA corpus with both unimodal and multimodal annotations. However, the time-consuming unimodal annotations result in a relatively small dataset size, further limiting the diversity of acoustic and visual emotion-bearing behaviors. As a result, we extend the CH-SIMS v2.0 based on CH-SIMS dataset with more expressive non-verbal behaviors in unlabeled instances to rich the multimodal context with a few workloads.

\subsection{Multimodal Sentiment Analysis}
\label{sec: rel_model}
 Previous MSA benchmarks mainly emphasize joint representation learning and multimodal fusion. For joint representation learning, \citet{wang2019words} construct a recurrent attended variation embedding network to shift textual representation according to the non-verbal cues. \citet{misa} present modality-invariant and modality-specific representations as joint multimodal representation. For multimodal fusion, \citet{tfn} propose a tensor fusion network, which obtains a new tensor representation by computing the outer product between unimodal representations. 
%  \citet{lmf} improve previous tensor fusion method through weight tensor decomposition and decrease the computational complexity. Recently delicate attention mechanisms are proposed for cross-modal fusion. 
 \citet{mfn} design a memory fusion network for cross-view interactions. \citet{mult} propose cross-modal transformers, which learn the cross-modal attention to reinforce a target modality. However, all previous MSA approaches assume that textual modality is dominant and fail to make full use of non-verbal behaviors.

\subsection{Mixup}
\label{sec: rel_mix}
% Mixup is a simple and effective data enhancement method, which can combine instances from the original data set to generate many virtual instances to improve the training volume of the model. 
Mixup is first proposed in Computer Vision (CV) as an effective regularization means to improve the generalization ability of the neural networks \cite{zhang2017mixup}. Based on the image feature Mixup, \citet{2021Interpolation} propose an Interpolation Consistency Training (ICT) method adapting the Mixup into the semi-supervised learning paradigm. 
% Due to its great power of improving model generalization ability, various Mixup strategy are designed for the natural language processing tasks. 
It can be categorized into input-level Mixup \cite{0CutMix, 2020SaliencyMix} and hidden-level Mixup \cite{2018Manifold} depending on the location of the mix operation. In literature \cite{2019Augmenting, 2020MixText}, they apply Mixup on hidden vectors like embeddings or intermediate representations. While \citet{2021SSMix} propose an input-level span wise Mixup method considering the salience of spans for text classification. \citet{liesting2021data} propose a word-level Mixup strategy for text Aspect-Based Sentiment Analysis. On this basis, we transfer it to the multimodal sentiment analysis task. In this work, we design a multimodal input-level Mixup strategy to enhance the acoustic and visual modality representation learning based on multitask late fusion backbone for MSA task.

\begin{figure}[htbp]
  \centering
  \includegraphics[width=1.0\textwidth]{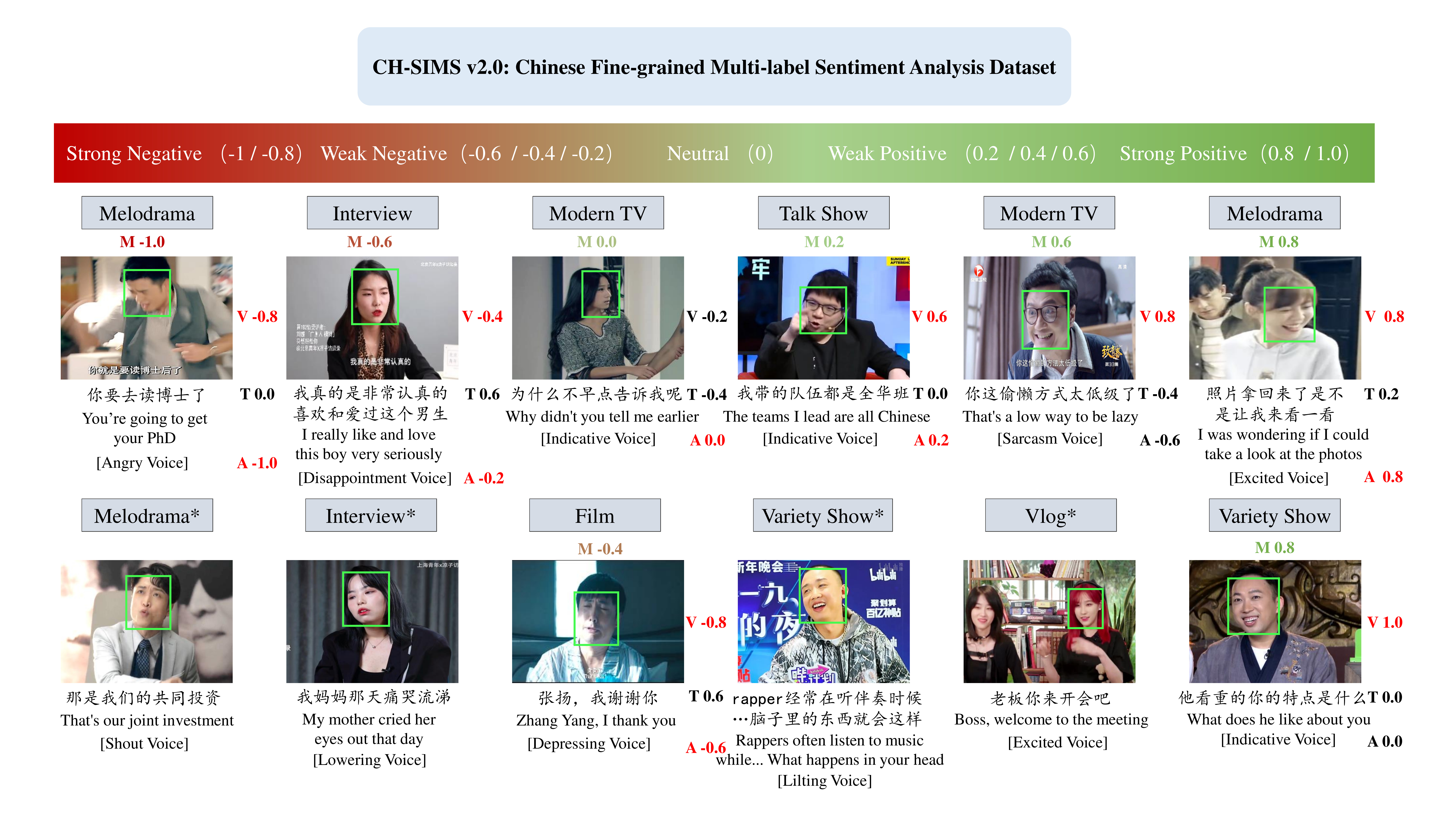}
  \caption{Illustration of the constructed CH-SIMS v2.0 dataset. The color of the multimodal annotation reflects the sentiment intensity from red (strong negative) to green (strong positive), while the color of the unimodal annotations on the right reflects the consistence between the unimodal and multimodal sentiment. Instances marked with "*" are unsupervised data.}
  \label{fig: CH-SIMS}
\end{figure}

\section{CH-SIMS v2.0 Dataset}
\subsection{Data Collection}
\label{sec: data_col}

Following the previous work, the raw video collection is conducted following the constraints that, for visual modality, high video definition with the speaker present in the video is required, for acoustic modality, the video should be in Mandarin, and for textual modality, accurate accompanying transcription is needed. The raw video segments are kept in their original resolution and recorded in MP4 format. After obtaining the raw videos, “PlotPlayer”\footnote{https://potplayer.daum.net/}, a popular video editing tool, is utilized to crop videos into segments at the frame level for high quality instances. Compared to the CH-SIMS dataset, the improvement of the video collection is reflected in two aspects.

\noindent\textbf{Diverse video scenarios.} It is natural that video segments in different scenes may have different sentiment tendencies. The previous CH-SIMS dataset \cite{yu2020ch} contains limited scenario resources due to the drawbacks of the traditional face detection toolkit, such as MTCNN \cite{zhang2016joint}, which can not distinguish the speaker's face from the multi-party scenarios. In our work, TalkNet \cite{tao2021someone} is utilized as the Active Speaker Detection (ASD) tool that enables the collection of video scenarios containing multiple faces. Under such circumstances, The CH-SIMS v2.0 dataset enriches its scenario diversity with raw video segments from melodrama, interviews, modern TV, talk show, vlogs, films, costume TV, variety show and many other scenes\footnote{Videos under creative commons license can be edited and used without the need for the authors' consent}. All video resolution is greater than 720p for better video quality from Bilibili, YouTube and television websites. To simulate complex real-world scenarios, the angle, distance from the camera, and light conditions might vary among different videos. The instances with missing speaker's faces will be removed for the instance quality, and instances with front, side and oblique faces will be collected. Moreover, the acoustic of the instance may contain slight noise, such as ambient music, and white noise, and the speaker's tone and speed might vary.

\noindent\textbf{Expressive acoustic and visual behavior.} Compared to the CH-SIMS, the proposed dataset focus more on non-verbal behaviors. Thus, instances of emotion-bearing non-verbal behaviors from the original cut video clips are consciously screened out. As shown in Figure \ref{fig: CH-SIMS}, for the instance \textbf{\textsc{"What does he like about you"}}, the neutral textual modality makes it important to capture the smiling face from visual modality. For the instance \textbf{\textsc{"I really like and love this boy very seriously”}}, textual modality is misleading for predicting the speaker's sentiment polarity. Moreover, there are other instances with ambiguity, ironies and metaphors text modality. In summary, CH-SIMS v2.0 contains plenty of instances with weak text modality reliance. Thus, text-based or text-predominant models might fail to predict the sentiment of the speakers.

\begin{figure}
  \centering
  \includegraphics[width=0.96\textwidth]{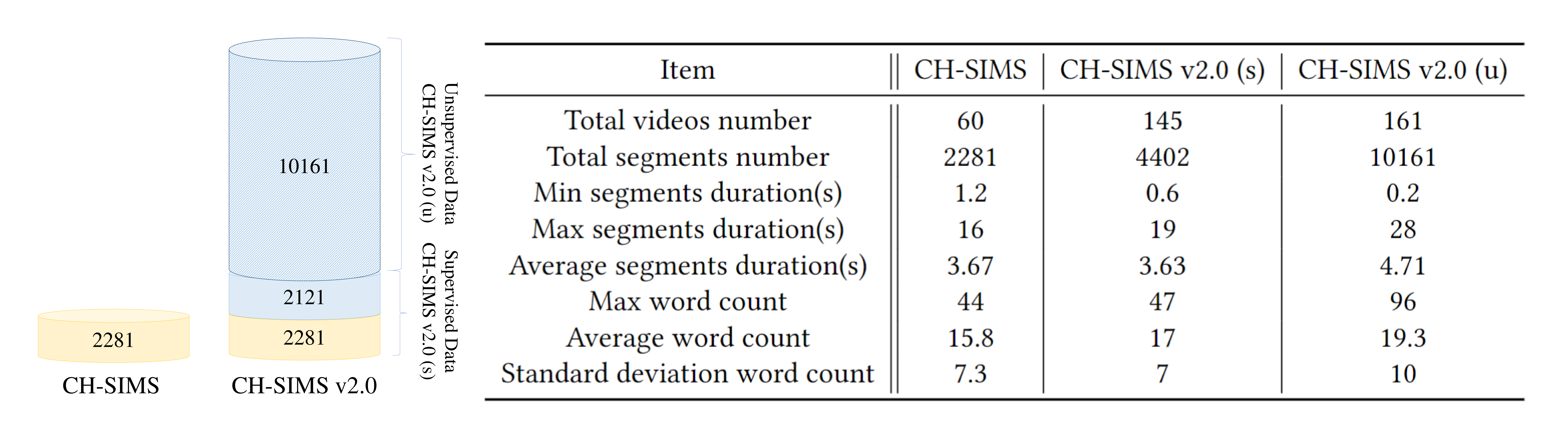}
  \caption{Statistics of each component of the CH-SIMS v2.0 dataset including the CH-SIMS v2.0 (s), and the CH-SIMS v2.0 (u). Instances in the original CH-SIMS dataset are relabeled and contained in the CH-SIMS v2.0 (s).}
  \label{fig: Total}
\end{figure}

The final statistics of the CH-SIMS v2.0 is shown in Fig \ref{fig: Total}. Specifically, the CH-SIMS v2.0 contains 4402 supervised instances, denoted as CH-SIMS v2.0 (s), and 10161 unsupervised instances, denoted as CH-SIMS v2.0 (u). For supervised resources, the CH-SIMS v2.0 (s) shares similar properties with the original dataset. While, for unsupervised resources, the CH-SIMS v2.0 (u) shows a much more diverse duration distribution simulating the potential real-world scenarios. In addition, the textual modality of the unsupervised instances is directly collected from the corresponding transcript without being manually refined and thus might contain noise.

\subsection{Data Annotation}
\label{sec: data_ann}
With the collection of raw video segments, data annotation on the supervised CH-SIMS v2.0 (s) is conducted. Consistent with the previous CH-SIMS dataset, each instance is annotated with fine-grained sentiment unimodal and multimodal labels ranging from Strong Negative (-3), Negative (-2), Weak Negative (-1), Neutral (0), Weak Positive (1), Positive (2), and Strong Positive (3). A total of seven annotators participate in the data labelling. Before labelling, several instances of each sentiment class are provided to annotators as the standard to improve fairness. In the post-processing period, the highest score and the lowest score are removed. The average score of the rest five results is then mapped to the space of Strong Negative (-1, -0.8), Weak Negative (-0.6, -0.4, -0.2), Neutral (0), Weak Positive (0.2, 0.4, 0.6), and Strong Positive (0.8, 1.0) as the final sentiment label. 

It is worth noticing that, CH-SIMS v2.0 improves the unimodal label annotation process with a strict modality isolation strategy and relabel the unimodal annotations of the original CH-SIMS dataset. Specifically, for acoustic modality labelling, noise is injected to blur the spoken word while maintaining the overall rhythm and acoustic features, for visual modality labelling, videos are presented in silent mode with the transcripts removed, and for textual modality labelling, only transcripts are provided. Moreover, to reduce the influence of inter-modal information on the annotator, annotation is carried out in the order of text, acoustic, visual and multimodal.

\begin{figure}[htbp]
  \centering
  \includegraphics[width=0.85\textwidth]{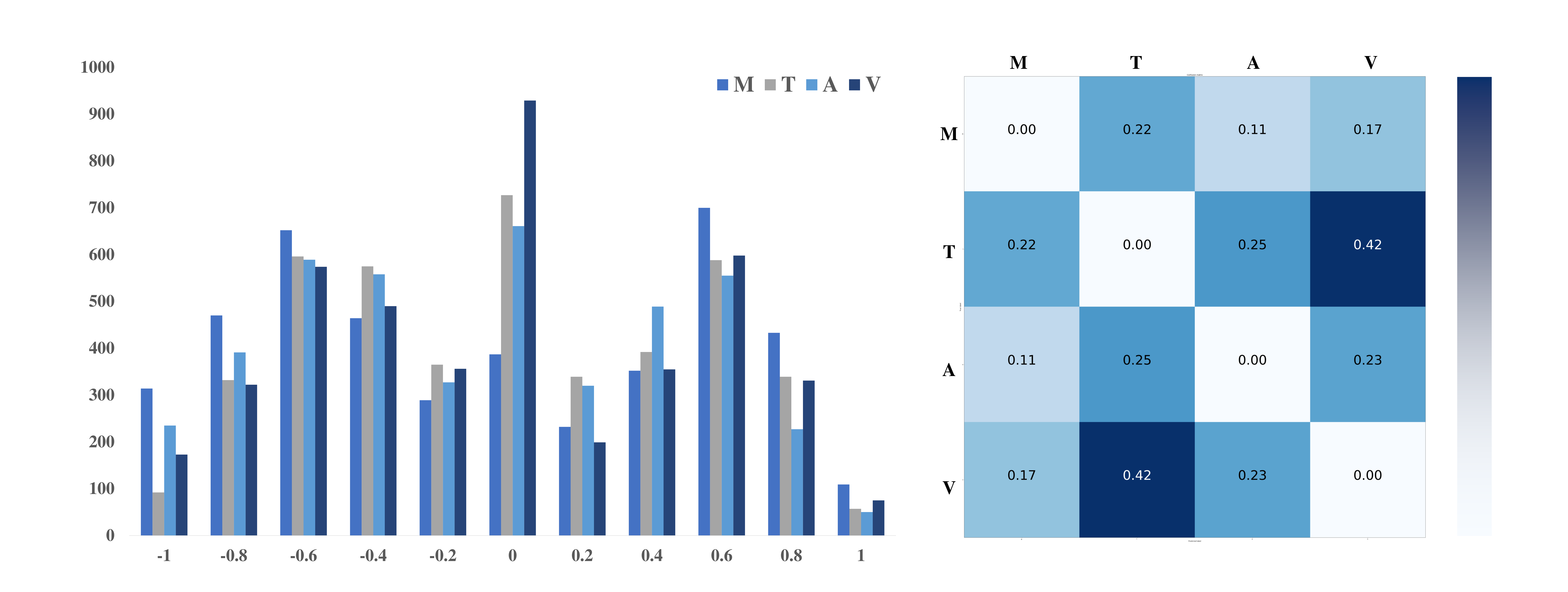}
  \caption{Left: the distribution of sentiment over the entire dataset in one Multimodal annotation and three unimodal (Text, Acoustic, and Visual) annotations. Right: the confusion matrix shows the annotations difference between different modalities in CH-SIMS (s) v2.0. The larger the value, the greater the difference.}
  \label{fig: Metric}
\end{figure}

The annotation statistics are shown on the left of Figure \ref{fig: Metric}. It can be found that in the CH-SIMS v2.0 dataset, the ratio of positive to negative instances is 83.41\% (56.38\% in the original CH-SIMS), and most instance position in the weak sentiment intensity scope. Besides, from the comparison of neutral instance count among different unimodal and multimodal annotations, it can be summarized that multimodal contains much more sentiment than each unimodal especially the visual modality revealing the importance of integrating multimodal information. In addition, following the literature \cite{yu2020ch}, the average discrepancy among each unimodal and multimodal label is illustrated on the right of the Fig \ref{fig: Metric}. Significant discrepancy validates the assumption that the unified multimodal labels can not reflect the unimodal sentiments. Furthermore, we can find that acoustic is the most similar to multimodal annotations, while the textual annotation is the least. All the above observations show the great potential to improve the sentiment prediction performance by making full use of the non-verbal cues.
% \begin{equation}
% D_{i j}=\frac{1}{N} \sum_{n=1}^{N}\left(A_{i}^{n}-A_{j}^{n}\right)^{2}
% \end{equation}
% where $i, j \in\{m, t, a, v\}$, $\mathrm{N}$ is the number of all instances, $A_{i}^{n}$ means the $n_{t h}$ label value in model $\dot{i}$. There are obvious differences in sentiment values among different modalities, and we can find that acoustic modality is close to multimodal, and language modality is the most different from multimodal. 
\subsection{Feature Extraction}
\label{sec: data_fea}

The default modality sequences for the CH-SIMS v2.0 are extracted through MMSA-FET, an open-sourced integrated feature extraction toolkit \cite{mao2022m}. In the following experiments, without additional description, default features are used.

\noindent\textbf{Default Textual Feature.} A pretrained BERT \cite{devlin2018bert} model, bert-base-chinese\footnote{https://huggingface.co/bert-base-chinese}, is used to learn contextual word embeddings as effective textual features. The length of token sequences is fixed to 50, using either padding or truncating method. The final textual feature is a word vector sequence in 768 dimensions.

\noindent\textbf{Default Acoustic Feature.} For acoustic features, 25-dimensional eGeMAPS \cite{egemaps} Low Level Descriptors (LLD) features are extracted by OpenSMILE \cite{eyben2010opensmile} backend at 16000 Hz sampling rate. The final acoustic features are padded or truncated to a sequence length of 925.

\noindent\textbf{Default Visual Feature.} For visual modality, images are first extracted using FFmpeg at 25 frames per second. TalkNet \cite{tao2021someone}, an effective Action Speaker Detection (ASD) method, is then used to detect the speaker’s face among all faces in a single image. The images over which the ASD failed are dropped, and instances with over 25\% missing images are discarded. After ASD, the OpenFace \cite{eyben2010opensmile} backend is used to extract the facial features, including 68 facial landmarks, 17 facial action units, head pose, head orientation, and eye gaze direction. Finally, the 177-dimensional frame-level visual features are padded or truncated to a sequence length of 232.

\section{Acoustic Visual Mixup Consistent (AV-MC) Framework}
\begin{figure}
  \centering
  \includegraphics[width=0.92\textwidth]{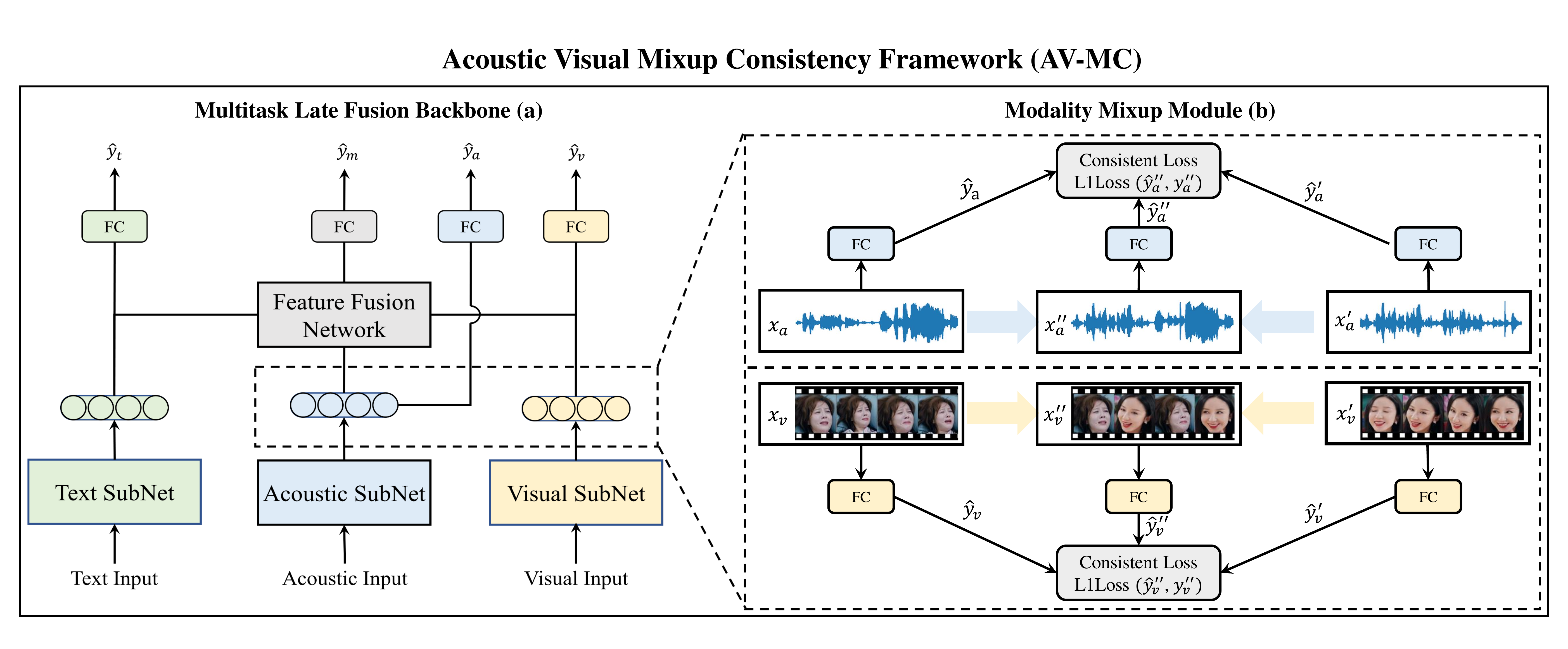}
  \caption{Acoustic Visual Mixup Consistent (AV-MC) Framework under semi-supervised learning paradigm, which consists of Multitask Late Fusion Backbone (a) with Modality Mixup Module (b).}
  \label{fig: AV-MC}
\end{figure}

Although the CH-SIMS v2.0 dataset strives to provide as many multimodal scenarios as possible, the potential multimodal context is not exhaustive. To further prepare the MSA model with unobserved multimodal context, we design the Acoustic Visual Mixup Consistent (AV-MC) framework. The overall structure of the AV-MC is illustrated in Figure \ref{fig: AV-MC}. 

\subsection{Modality Mixup Module}
\label{sec: model_str}
The modality mixup module aims to generate potential modality representation with corresponding annotations from the original representation obtained from the modality encoder. Specifically, given the set of instances with corresponding annotations $\{(\mathbf{X}_1, y_1), \cdots ,(\mathbf{X}_n, y_n)\}$, a random shuffle is first conducted,
\begin{equation}
\{(\mathbf{X}^{\prime}_1, y^{\prime}_1), \cdots, (\mathbf{X}^{\prime}_n, y^{\prime}_n)\} = \text{Shuffle}( \{(\mathbf{X}_1, y_1), \cdots, (\mathbf{X}_n, y_n)\}).
\end{equation}
After shuffling, the mixup is performed on the shuffled and original instance pair through the weighted average,

\begin{equation}\label{formula: XMix}
{X}^{\prime\prime}_{i}=\lambda \cdot X_{i}+(1-\lambda) \cdot {X}^{\prime}_{i},
\end{equation}
\begin{equation}\label{formula: YMix}
y^{\prime\prime}_{i}=\lambda \cdot y_{i}+(1-\lambda) \cdot {y}^{\prime}_{i},
\end{equation}                                                                                                                                                                                           
where $\lambda \in [0,1]$ is a random variable sampled from $\textbf{Beta}$ distribution. Using the above definition, the modality mixup module is formulated as follows,

\begin{equation}
    \{(\mathbf{X}^{\prime\prime}_1, y^{\prime\prime}_1), \cdots ,(\mathbf{X}^{\prime\prime}_n, y^{\prime\prime}_n)\} = \textbf{Mixup}(\{(\mathbf{X}_1, y_1), \cdots ,(\mathbf{X}_n, y_n)\}).
\end{equation}
The generated $\{(\mathbf{X}^{\prime\prime}_1, y^{\prime\prime}_1), \cdots ,(\mathbf{X}^{\prime\prime}_n, y^{\prime\prime}_n)\} $ can be used as potential instances promoting representation learning.

\subsection{Supervised and Semi-supervised learning with AV-MC Framework}
\label{sec: model_tra}

% The proposed AV-MC framework utilizes trivial multitask late fusion as backbone.
For the convenience in the following introduction, the initial \textbf{t}ext, \textbf{a}udio and \textbf{v}ision sequences are denoted as $\mathbf{I}_k \in \mathcal{R}^{T_k\times d_k}$, where $T_k$ refers to the modality sequence length, and $d_k$ refers to the initial feature dimension ($k\in \{t, a, v\}$). Receiving the initial modality sequences as input, the unimodal encoders are first utilized for both supervised and unsupervised instances to learn inter-modal representations. The unified unimodal encoders are formulated as follows.

% \noindent\textbf{Unimodal Encoders.} The unified unimodal encoder is used to transform the initial modality sequence into dense semantic vector,
\begin{equation}\label{formula: Extractor}
\mathbf{F}_k = \mathbf{S}_k\left(\mathbf{I}_k\right) \in \mathcal{R}^{h_k}, 
\end{equation}
where $k\in \{t, a, v\}$, $h_k$ is the hidden dimension for modality $u$, and $\mathbf{S}_u(\cdot)$ represents the unimodal encoder network. Specifically, in the proposed AV-MC framework, for textual modality, a one layer feed-forward network is regarded as the encoder transforming the first time step vector, which refers to [CLS]\footnote{a special token in Bert language model, appended at the front of the token sequence, commonly used for sentence-level representation learning} token into the textural representation. For acoustic and visual modalities, a stack bi-directional Long Short Term Memory (LSTM) \cite{1997Long} followed by a one layer feed-forward network is utilized as the encoder. The one layer feed-forward network is defined as,
\begin{equation}
\begin{aligned}
\text{FFN}(\mathbf{x}) = \sigma(\mathbf{W} \cdot \mathbf{x}+ \mathbf{b}),
\end{aligned}
\end{equation}
where $\sigma$ represents the activation function, $\mathbf{W}$ and $\mathbf{b}$ are learnable model parameters. As a simple but effective fusion strategy, we take the concatenation of the unimodal representation as to the final multimodal representation.

\begin{equation}\label{formula: cat}
\mathbf{F}_m=\text {Concat}\left(\left[\mathbf{F}_t ; \mathbf{F}_a ; \mathbf{F}_v\right]\right) \in \mathcal{R}^{h_t + h_a + h_v},
\end{equation}
After obtaining the unimodal and the multimodal representations, for supervised instances, four independent classifiers consisting of a three layers feed-forward network are utilized for both unimodal and multimodal sentiment prediction,

\begin{equation}\label{formula: FC}
\hat{y}_{k} = \text{Clf}_{k}(\mathbf{F}_k) = \text{FFN}(\text{FFN}(\text{FFN}(\text{BN}(\mathbf{F}_k)))) \in \mathcal{R},
\end{equation}
where $k \in \{m, t, a, v\}$, and $\text{BN}$ refers to the batch normalization. $\text{L1\_Loss}$ is used as supervision for each unimodal and multimodal task,
\begin{equation}\label{formula: clfLoss}
\mathcal{L}_{r}^{(k)} = \frac{1}{N_s}\sum_{i=1}^{N_s}|\hat{y}_{k}^{(i)} - y_{k}^{(i)}|, 
\end{equation}
where $k \in \{m, t, a, v\}$, $N_s$ is the supervised instances count. The final sentiment regression loss is formulated as the weighted average of the unimodal and multimodal tasks,
\begin{equation}
\label{formula: clfLossAll}
\mathcal{L}_{r} = \sum_k \alpha_k \cdot \mathcal{L}_{r}^{(k)}
\end{equation}
where $\alpha_k, k\in \{m, t, a, v\}$ is the hyper-parameter balancing the contribution of unimodal and multimodal tasks. In addition to the regression tasks for the supervised instances, mixup consistent tasks are performed for both supervised and unsupervised data on acoustic and visual modalities. The acoustic and visual representation $\mathbf{F}_k$ along with corresponding prediction $\hat{y}_k$ is passed through the modality mixup module,
\begin{equation}\label{formula: Mixup}
\left\{(\mathbf{F}_k^{\prime(1)}, \hat{y}_k^{\prime(1)}), \cdots , (\mathbf{F}_k^{\prime(N_s+N_u)},\hat{y}_k^{\prime(N_s+N_u)})\right\} = \textbf{Mixup}\left(\left\{(\mathbf{F}_k^{(1)}, \hat{y}_k^{(1)}), \cdots , (\mathbf{F}_k^{(N_s+N_u)},\hat{y}_k^{(N_s+N_u)})\right\}\right),
\end{equation}
where $k\in\{a, v\}$, and $N_s, N_u$ refers to the total instances count for CH-SIMS v2.0 (s) and CH-SIMS v2.0 (u). With the generated acoustic unimodal and visual unimodal representations, the same acoustic and visual unimodal classifiers are utilized for both mixed acoustic unimodal and mixed visual unimodal sentiment prediction. 

\begin{equation}\label{formula: MixFC}
\hat{y}_{k}^{\prime\prime} = \text{Clf}_{k}(\mathbf{F}_k) = \text{FFN}(\text{FFN}(\text{FFN}(\text{BN}(\mathbf{F}_k)))) \in \mathcal{R}, 
\end{equation}
where $\hat{y}_{k}^{\prime\prime}$ is the model prediction for the generated instance $\mathbf{F}_k^{\prime}$. Inspired from literature \cite{verma2019interpolation}, the generated instances serve as the interpolation of the two original instances, the prediction of which should be consistent with the direct average with the same weights, i.e. $\hat{y}_{k}^{\prime}$. We still use $\text{L1\_Loss}$ for mixup consistent tasks.
\begin{equation}\label{formula: consistentLoss}
\mathcal{L}_{mix}^{(k)} = \frac{1}{N_s+N_u}\sum_{i=1}^{N_s+N_u}|\hat{y}_{k}^{\prime\prime(i)} - \hat{y}_{k}^{\prime(i)}|, 
\end{equation}
where $k \in \{a, v\}$. The final consistent loss is formulated as the weighted average of the acoustic and visual mixup tasks,
\begin{equation}
\label{formula: consistentLossAll}
\mathcal{L}_{mix} = \sum_k \beta_k \cdot \mathcal{L}_{mix}^{(k)}
\end{equation}
where $\beta_k, k\in \{a, v\}$ is the hyper-parameter balancing the contribution of acoustic and visual mixup tasks. 

Each epoch in the training process contains two iterations. The first iteration uses both supervised and unsupervised instances and updates the model parameters under the guidance of both the regression loss $\mathcal{L}_r$ and the mixup consistent loss $\mathcal{L}_{mix}$. To promote the model convergence, an additional iteration on the supervised instances is conducted using only the regression loss for supervision. The detailed training process is shown in the algorithm (\ref{alg: semi-supervised}). 
\begin{algorithm}
\small
    \renewcommand{\algorithmicrequire}{\textbf{Input:}}
    \caption{Semi-supervised Training Strategy}
    \label{alg: semi-supervised}
    \begin{algorithmic}[1]
        \REQUIRE unimodal inputs $\mathbf{I}_t$, $\mathbf{I}_a$, $\mathbf{I}_v$, labels $y_{c}$, unimodal labels $y_t y_a y_v$, multimodal labels $y_m$ 
        \STATE Initialize model parameters
        \FOR {$n \in [1, end]$}
            \FOR {$\text{mini-batch}\in \text{dataloader (supervised data and unsupervised data)}$}
            % \TCC{Calculate Grad for TAVM Module.}
                \STATE $\mathbf{R}_{t},\mathbf{R}_{a},\mathbf{R}_{v} \gets \mathbf{S}_t(\mathbf{I}_t), \mathbf{S}_a(\mathbf{I}_a), \mathbf{S}_v(\mathbf{I}_v)$ using Equation \ref{formula: Extractor}
                \STATE $\mathbf{R}_{m} \gets$ Concat ([$\mathbf{R}_{t}, \mathbf{{R}_{a}}, \mathbf{R}_{v}$]) using Equation \ref{formula: cat}
                \STATE $\hat{y}_t, \hat{y}_a, \hat{y}_v, \hat{y}_m \gets \mathbf{Clf}_t(\mathbf{R}_t), \mathbf{Clf}_a(\mathbf{R}_a), \mathbf{Clf}_v(\mathbf{R}_v), \mathbf{Clf}_m(\mathbf{R}_m)$ using Equation \ref{formula: FC}
                % \STATE Compute gradients for t, a, v, m classification (Only for supervised data.)
                \STATE Compute t, a, v, m sentiment regression loss        $\mathcal{L}^{(t)}_{r}, \mathcal{L}^{(a)}_{r}, \mathcal{L}^{(v)}_{r}, \mathcal{L}^{(m)}_{r}$ using Equation \ref{formula: clfLoss}.
                \STATE Compute final sentiment regression loss $\mathcal{L}_{r}$ using Equation \ref{formula: clfLossAll}
                \STATE $\mathbf{R}^{\prime}_{a}, \hat{y}_a^{\prime} \gets \mathbf{Mixup}(\mathbf{R}_a, \hat{y}_a)$ using Equation \ref{formula: Mixup}
                \STATE $\hat{y}_{a}^{\prime\prime} \gets \mathbf{Clf}_a(\mathbf{R}^{\prime}_{a})$ using Equation \ref{formula: MixFC}
                \STATE Compute acoustic mixup consistent Loss $\mathcal{L}^{(a)}_{mix}$ using Equation \ref{formula: consistentLoss}
                \STATE $\mathbf{R}^{\prime}_v, \hat{y}_v^{\prime} \gets \mathbf{Mixup}(\mathbf{R}_v, \hat{y}_v)$ using Equation \ref{formula: Mixup}
                \STATE $\hat{y}_{v}^{\prime\prime} \gets \mathbf{Clf}_v(\mathbf{R}^{\prime}_{v})$ using Equation \ref{formula: MixFC}
                \STATE Compute visual mixup consistent Loss $\mathcal{L}^{(v)}_{mix}$ using Equation \ref{formula: consistentLoss}
                \STATE Compute final consistent loss $\mathcal{L}_{mix}$ using Equation \ref{formula: consistentLossAll}
                \STATE Update model parameters.
            \ENDFOR
            \FOR {$\text{mini-batch}\in \text{dataloader(supervised data)}$}
                \STATE $\mathbf{R}_{t},\mathbf{R}_{a},\mathbf{R}_{v} \gets \mathbf{S}_t(\mathbf{I}_t), \mathbf{S}_a(\mathbf{I}_a), \mathbf{S}_v(\mathbf{I}_v)$ using Equation \ref{formula: Extractor}
                \STATE $\mathbf{R}_{m} \gets$ Concat ([$\mathbf{R}_{t}, \mathbf{{R}_{a}}, \mathbf{R}_{v}$]) using Equation \ref{formula: cat}
                \STATE $\hat{y}_t, \hat{y}_a, \hat{y}_v, \hat{y}_m \gets \mathbf{Clf}_t(\mathbf{R}_t), \mathbf{Clf}_a(\mathbf{R}_a), \mathbf{Clf}_v(\mathbf{R}_v), \mathbf{Clf}_m(\mathbf{R}_m)$ using Equation \ref{formula: FC}
                % \STATE Compute gradients for t, a, v, m classification (Only for supervised data.)
                \STATE Compute t, a, v, m sentiment regression loss $\mathcal{L}^{(t)}_{r}, \mathcal{L}^{(a)}_{r}, \mathcal{L}^{(v)}_{r}, \mathcal{L}^{(m)}_{r}$ using Equation \ref{formula: clfLoss}.
                \STATE Compute final sentiment regression loss $\mathcal{L}_{r}$ using Equation \ref{formula: clfLossAll}
                \STATE Update model parameters.
            \ENDFOR
        \ENDFOR
    \end{algorithmic}
\end{algorithm}

\section{Experiments and Discussion}

% This section presents the experimental result and discussion on the CH-SIMS v2.0 starting with the evaluation metrics in Section \ref{sec: exp_met}. Benchmark MSA model performances on the CH-SIMS v2.0 are recorded in Section \ref{sec: exp_bas}. Case study on CH-SIMS v2.0 is conducted and demonstrated in Section \ref{sec: exp_cas}. In addition to the constructed CH-SIMS v2.0, the effectiveness of the proposed AV-MC framework is further validated through ablation study in Section \ref{sec: exp_abl}.
\begin{table}[t]
\small
\centering
\begin{tabular}{c|c|ccccc}
\toprule[1pt]
Item & Total & NEG & WNEG & NEU & WPOS & POS \\
\midrule[1pt]
\#Train & 2722 & 921 & 433 & 232 & 318 & 818\\
\#Valid & 647  & 224 & 110 & 62 & 83 & 168\\
\#Test & 1034  & 291 & 211 & 93 & 183 & 256\\
\bottomrule[1pt]
\end{tabular}
\caption{Data splits in CH-SIMS v2.0 (s). NEG: Negative, WNEG: Weak Negative, NEU: Neutral, WPOS: Weak Positive, POS: Positive.}
\label{res: Data partitioning}
\end{table}

\subsection{Dataset splits and Evaluation Metric}
\label{sec: exp_met}
As presented in Table \ref{res: Data partitioning}, the CH-SIMS v2.0 dataset is partitioned into training, validation and test sets in a ratio close to 9:2:3. Following the previous work, the model performances are evaluated in the form of classification and regression. Traditional classification metrics including binary classification accuracy (Acc2) and F1 score (F1\_Score) reflect the correctness of basic sentiment polarity prediction, i.e. positive or negative classification. In addition, Acc2\_weak is utilized to further validate the model performance for the weak emotion instances labelled in [-0.4, 0.4]. Traditional regression metrics including Mean Absolute Error (MAE) and the Pearson Correlation (Corr) are recorded for fine-grained prediction evaluation. 
% formulated in Equation \ref{formula: Corr},
% \begin{equation}
% \begin{aligned}
%     \text{MAE}(Y,\hat{Y}) &= \frac{1}{n} \sum_{i=1}^{n} |y_i - \hat{y}_i|,\\
%     \text{Corr}(Y,\hat{Y}) &= \frac{\sum_{i=1}^{n}\left(y_{i}-\bar{y}\right)\left(\hat{y}_{i}-\bar{\hat{y}}\right)}{\sqrt{\sum_{i=1}^{n}\left(y_{i}-\bar{y}\right)^{2}} \sqrt{\sum_{i=1}^{n}\left(\hat{y}_{i}-\bar{\hat{y}}\right)^{2}}},
%     \label{formula: Corr}
% \end{aligned}
% \end{equation}
% where, $\bar{y} = \frac{1}{n}\sum_{i=1}^n y_i$ is the average of the ground truth. 
Moreover, the R\_square is also used to compare the model performance against the trivial solution (prediction of the average result on the test set).
% formulated in Equation \ref{formula: R},
% \begin{equation}\label{formula: R}
%  \text{R}^{2}(Y,\hat{Y})=\frac{\sum_{i=1}^{n}\left(\bar{y}-y_{i}\right)^{2}-\sum_{i=1}^{n}\left(\hat{y}_{i}-y_{i}\right)^{2}}{\sum_{i=1}^{n}\left(\bar{y}-y_{i}\right)^{2}},
% \end{equation}
% where, $\bar{y} = \frac{1}{n}\sum_{i=1}^n y_i$ is the average of the ground truth. 
For all the above metrics, higher values indicate better model performance, except MAE, where lower values indicate better model performance.

\subsection{Benchmark Results on CH-SIMS v2.0}
\label{sec: exp_bas}

% To facilitate further research using the CH-SIMS v2.0, the model performance of thirteen MSA benchmarks are provided. The selected benchmarks is introduced briefly.

\begin{table*}[t]
\small
\centering
\begin{tabular}{c||cccccc}
\toprule[1pt]
Models & Acc2 ($\uparrow$) & F1\_score ($\uparrow$) & Acc2\_weak ($\uparrow$) & Corr ($\uparrow$) & R\_squre ($\uparrow$) & MAE ($\downarrow$)\\
% \midrule[1pt]
% Text & 75.01 & 74.98 & 66.21 & 59.49 & 32.03 & 0.371\\
% Acoustic & 57.16 & 57.29 & 54.00 & 18.08 & -5.09 & 0.484\\
% Visual & 71.76 & 71.78 & 65.30 & 50.35 & 18.51 & 0.402\\
\midrule[1pt]
LF\_DNN     & 73.95 & 73.84 & 69.13 & 52.19 & 20.84 & 0.381 \\
TFN        & 76.51 & 76.31 & 66.27 & 66.65 & 35.90 & 0.323 \\
LMF        & 77.05 & 77.02 & 69.34 & 63.75 & 40.64 & 0.343 \\
MFN        & 75.27 & 75.24 & 66.46 & 60.60 & 32.26 & 0.355 \\
Graph\_MFN & 73.98 & 73.62 & 69.82 & 49.71 & 13.78 & 0.396 \\
MulT       & 79.50 & 79.59 & 69.61 & 70.32 & 47.15 & 0.317 \\
Bert\_MAG   & 79.79 & 79.78 & 71.87 & 69.09 & 43.08 & 0.334 \\
MISA       &  80.53 & 80.63 & 70.50 & 72.49 & 50.59 & 0.314 \\
MMIM       &  80.95 & 80.97 & 72.28 & 70.65 & 43.81 & 0.316 \\
Self\_MM       & 79.01 & 78.89 & 71.87 & 64.03 & 29.36 & 0.335 \\
\midrule[1pt]
MLF\_DNN$^{*}$      & 78.40 & 78.44 & 71.59 & 65.80 & 39.34 & 0.326 \\
MTFN$^{*}$          & 80.26 & 80.33 & 71.07 & 70.54 & 46.07 & 0.318 \\
MLMF$^{*}$          & 79.92 & 79.72 & 69.88 & 71.37 & 47.53 & 0.302 \\
\midrule[1pt]
AV-MC$^{*}$         & 82.50 (2.00\%) & 82.55 (2.01\%) & 74.54 (3.13\%) & 73.17 (0.94\%) & 50.65 (0.12\%) & 0.297 (1.66\%) \\
\textbf{AV-MC(Semi)$^{*}$}   & \textbf{83.46 (3.10\%)} & \textbf{83.52 (3.15\%)} & \textbf{74.54 (3.13\%)} & \textbf{76.04 (4.90\%)} & \textbf{57.37 (13.40\%)} & \textbf{0.286 (5.30\%)} \\
\bottomrule[1pt]
\end{tabular}
\caption{Model performances for Traditional Multimodal Sentiment Analysis model on CH-SIMS v2.0 dataset. Models with \textbf{$^{*}$} are trained on multitasking, (Semi represents the additional use of unsupervised data). The best results are highlighted in bold.}
\label{res: main results}

\end{table*}
\subsubsection{Baselines}
\label{sec: baselines}
The benchmark models can be segmented into two classes according to whether unimodal annotations are used or not. The traditional MSA models which are supervised with the unified multimodal annotations include the Late Fusion Deep Neural Network (LF\_DNN) \cite{lf_dnn_ef_lstm}, the Tensor Fusion Network (TFN) \cite{tfn}, the Low-rank Multimodal Fusion (LMF) \cite{lmf}, the Memory Fusion Network (MFN) \cite{mfn}, the Graph Memory Fusion Network (Graph\_MFN) \cite{mosei}, the Multimodal Transformer (MulT) \cite{mult}, the Multimodal Adaptation Gate for Bert Network (Bert\_MAG) \cite{BERT_MAG}, the Modality-Invariant and -Specific Representations Network (MISA) \cite{misa}, the Multimodal InfoMax Network (MMIM) \cite{MMIM}, the Self-Supervised multitask Learning Network (Self\_MM) \cite{self-mm}. The others which utilize the unimodal annotations to guide unimodal representation learning include the Multi-task Tensor Fusion Network (MTFN), the Multitask Late Fusion Deep Neural Network (MLF\_DNN), the Multitask Low-rank Multimodal Fusion (MLMF).

\subsubsection{Benchmark Model Performances}
Table \ref{res: main results} presents the model performances on the CH-SIMS v2.0 dataset. Among different evaluation metrics, all benchmark models show relative low performance on Acc2\_weak and all regression metrics. Such a phenomenon reveals that the current MSA benchmarks fail in fine-grained sentiment intensity prediction, though they perform well in basic sentiment polarity classification. Moreover, from the comparison between two different class benchmarks (described in Section \ref{sec: baselines}), it can be found that taking advantage of the unimodal annotations models with a simple late fusion backbone can achieve competitive performances with state-of-the-art models. The convincing performance of the second class benchmarks further shows the effectiveness of the unimodal annotations for the fine-grained sentiment intensity prediction. Furthermore, the proposed AV-MC framework outperforms all existing MSA benchmarks, especially in Acc2\_weak under the supervised settings, validating the effectiveness of the modality mixup strategy for discriminating instances with weak emotion. Finally, the model performance is further improved with the CH-SIMS v2.0 (u) revealing the potential improvement with the unsupervised data.

\begin{table}[t]
\small
\centering
\begin{tabular}{cc||cccc}
\toprule[1pt]
Modality & Feature & Acc2 ($\uparrow$) & Acc2\_weak ($\uparrow$) & Corr ($\uparrow$) & MAE ($\downarrow$)\\
\midrule[1pt]
Text & default & 78.72 / 75.21 & 67.12 / 62.49 & 75.45 / 59.49 & 0.252 / 0.371 \\
Acoustic & default & 57.16 / 56.48 & 55.61 / 54.41 & 13.48 / 13.72 & 0.424 / 0.491 \\
Acoustic & wav2vec 2.0 & 65.38 / 60.93 & 59.96 / 53.01  & 36.71 / 38.05  & 0.396 / 0.446 \\
Visual & default & 78.72 / 73.11 & 76.04 / 68.79 & 57.70 / 48.54 & 0.314 / 0.401 \\
\bottomrule[1pt]
\end{tabular}
\caption{Model performances for unimodal sentiment analysis with both unimodal annotation and multimodal annotation. For each metric, model performances with unimodal annotation are shown on the left.}
\label{res: unimodal results}
\end{table}

In addition, unimodal sentiment analysis with both unimodal annotation and multimodal annotation is conducted. The former utilizes unimodal annotation for supervision and evaluation, while the latter uses multimodal annotation for supervision and evaluation. For all unimodal experiments, the same LSTM \cite{1997Long} based model is utilized for sentiment intensity regression. Experimental results are shown in Table \ref{res: unimodal results}. The relative bad performance on the acoustic unimodal task reveals the challenge of emotion-bearing acoustic feature extraction which is left in future research. Besides, the performance gap under unimodal and multimodal annotations verifies the assumption of multimodal annotation can be misleading for unimodal representation learning.

Furthermore, for the constructed CH-SIMS v2.0 dataset, the instance contains rich emotion-bearing non-verbal cues alleviating the text-predominant phenomenon. Moreover, the independent unimodal annotation along with the unsupervised instances further provides a potential solution for exploring the non-verbal context. The proposed AV-MC framework, which learns to be aware of non-verbal context through simple modality mixup, can be regarded as an initial attempt at the challenge of making full use of non-verbal behaviors. In the future, different feature combinations, and model frameworks can be designed on CH-SIMS v2.0 addressing on the above challenge.

\subsection{Case Study}
\label{sec: exp_cas}
\begin{figure}
  \centering
  \includegraphics[width=0.96\textwidth]{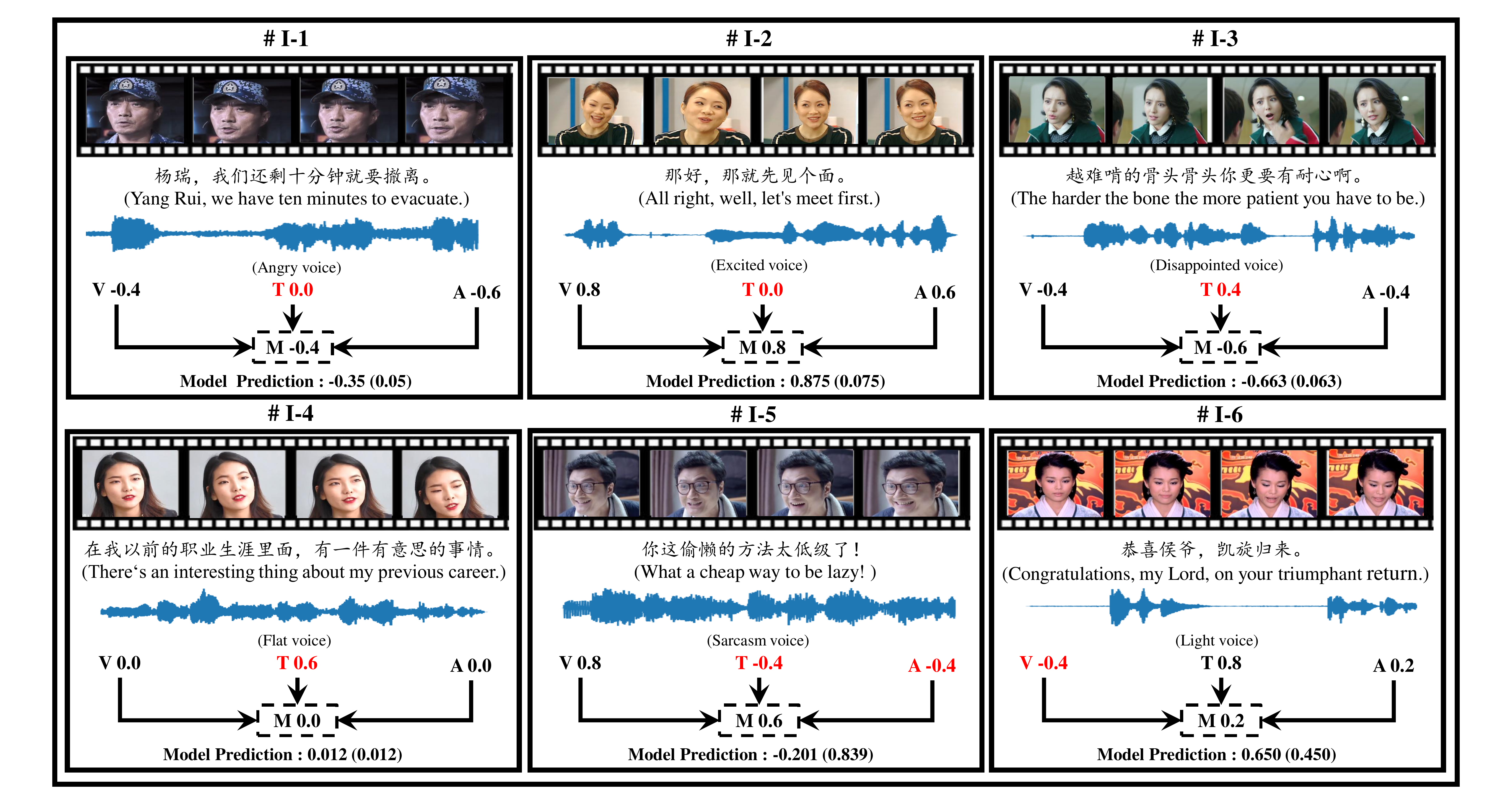}
  \caption{Case study results on CH-SIMS v2.0. For each instance, spoken words, non-verbal behaviors, unimodal annotations, and multimodal annotations are recorded where the inconsistent unimodal annotation is highlighted in red. The prediction of the proposed AV-MC framework is provided along with the MAE value.}
  \label{fig: case1}
\end{figure}
As shown in Figure \ref{fig: case1}, we illustrate some typical cases of the constructed CH-SIMS v2.0 with corresponding predictions and annotations. Firstly, intuitively, the cases show the diverse non-verbal emotion-bearing behaviors, such as the obvious angry, disappointed voice, and the smiling, stiff facial expression. Such rich non-verbal behaviors are crucial for fine-grained sentiment prediction and alleviate the text-predominant phenomenon. Secondly, from the model aspect, it can be observed that the proposed AV-MC framework performs well in the cases where more than two modality annotations are consistent with the ground truth multimodal annotations (\#I-1, \#I-2, \#I-3, \#I-4). Specifically, for the case \#I-2, the textual modality contains no obvious emotion. Nevertheless, the AV-MC framework still makes a prediction with less than 0.1 difference through the negative positive acoustic and visual cues. While for the case \#I-5, though the obvious smiling faces, the AC-MC still fail to predict the positive sentiment due to the negative expressions in the majority of modality (textual and acoustic). Above result reveals the proposed AV-MC module can improve performance by providing better unimodal representation learning. However, due to the limitation of the simple late fusion strategy, the AV-MC framework shows poor performance in sentiment inference. 

For robust real-world scenario applications, the CH-SIMS v2.0 provides resources to fully explore non-verbal behaviors. In the future, MSA models should contain effective unimodal encoders to capture the modality-specific features as well as a robust fusion module integrating the unimodal cues for the sentiment inference.
% \subsection{Unimodal Representation Visualization}

\begin{table*}[t]
\small
\centering
\begin{tabular}{c||cccccc}
\toprule[1pt]
 & Acc2 ($\uparrow$) & F1\_score ($\uparrow$) & Acc2\_weak ($\uparrow$) & Corr ($\uparrow$) & R\_squre ($\uparrow$) & MAE ($\downarrow$)\\
\midrule[1pt]
w/o Mixup-AV \& Unimodal tasks  & 78.72 & 78.71 & 70.43 & 67.49 & 40.16 & 0.334 \\
w/o Mixup-AV      & 80.47 & 80.38 & 71.66 & 73.08 & 50.32 & 0.321 \\
w/o Mixup-A      & 81.33 & 81.36 & 72.07 & 72.02 & 49.96 & 0.297 \\
w/o Mixup-V      & 80.17 & 80.17 & 71.46 & 71.73 & 48.28 & 0.305 \\
\midrule[1pt]
\textbf{AV-MC}     & \textbf{82.50} & \textbf{82.55} & \textbf{74.54} & \textbf{73.17} & \textbf{50.65} & \textbf{0.297} \\
\bottomrule[1pt]
\end{tabular}
\caption{Ablation study results for AV-MC on CH-SIMS v2.0 dataset. w/o Mixup-AV \& Unimodal tasks represents the removal of the entire Acoustic and Visual Mixup strategy and multitask training strategy; w/o Mixup-AV represents removal of both Acoustic and Visual Mixup strategy ; w/o Mixup-A and w/o Mixup-V represent removal of Acoustic Mixup strategy or Visual Mixup strategy in AV-MC framework separately. Best results are highlighted in bold.}
\label{res: Ablation1 results}
\end{table*}

\subsection{Ablation Study}
\label{sec: exp_abl}
Table \ref{res: Ablation1 results} presents the quantitative ablation study results. We first ablate the AV-MC framework by removing both acoustic visual mixup and the unimodal supervision in multitask late fusion backbone, denoted as w/o Mixup-AV \& Unimodal tasks. Under such circumstances, the model performance dramatically declines by 4.80\%, 5.84\%, 11.08\% on Acc2, Acc2\_weak and MAE separately. Based on the multitask framework with the supervision of unimodal and multimodal annotations, we then ablate the AV-MC framework by removing the entire mixup, acoustic mixup, and visual mixup strategies respectively denoted as w/o Mixup-AV, Mixup-A and Mixup-V. Through the performance comparison among the above three ablations and the original AV-MC framework, it can be found that both acoustic and visual modality mixup contribute to the overall model performance, especially for the visual mixup.
\begin{figure}[htbp]
   \centering
  \includegraphics[width=0.8\textwidth]{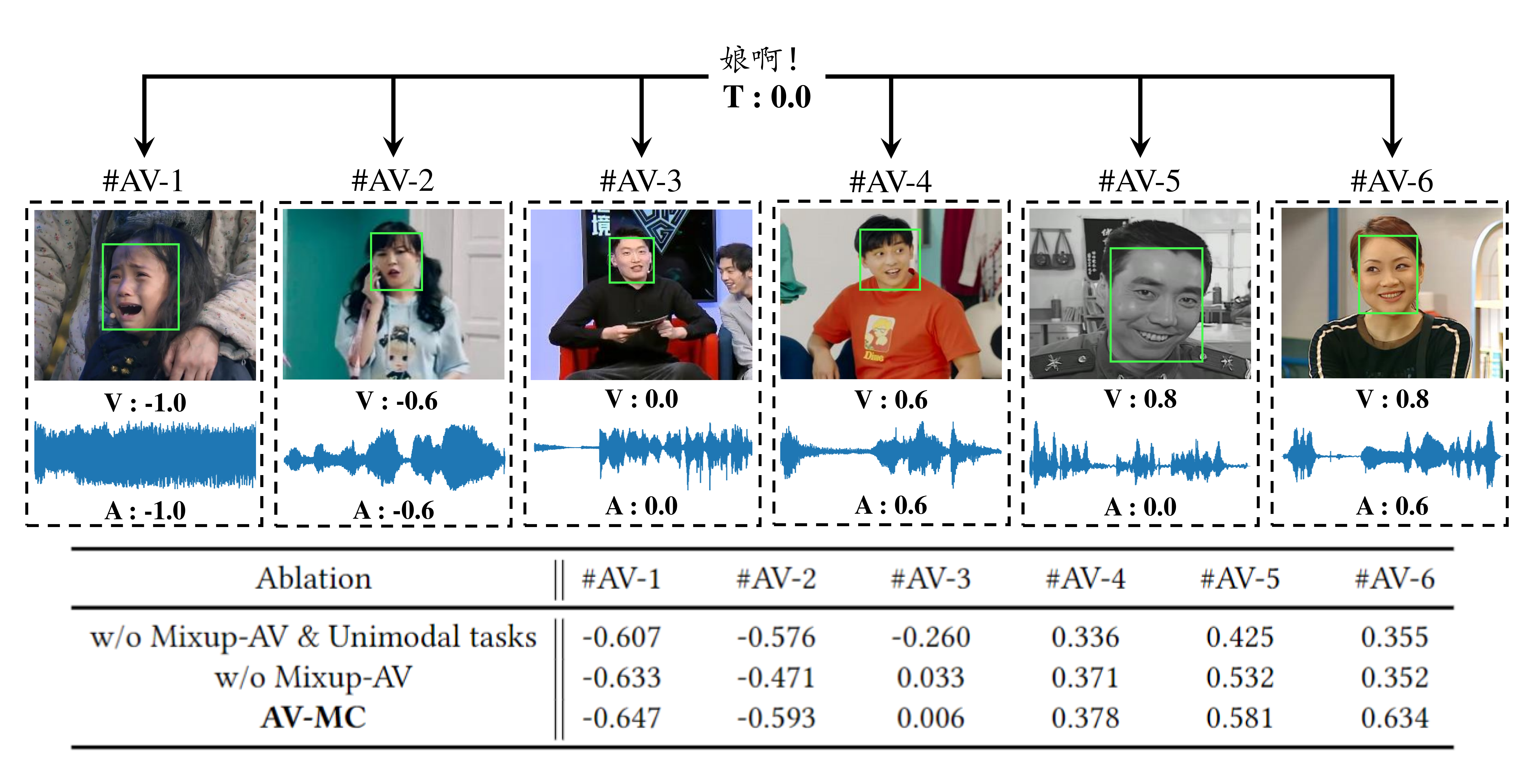}
  \caption{Ablation demonstration for the same spoken word combined with various expressive non-verbal scenarios.}
  \label{fig: ablate}
\end{figure}

To further explore how the modality mixup and the unimodal tasks help to improve the model performance, we manually construct six expressive non-verbal scenarios ranging from strongly negative to strongly positive for the same neutral spoken words \textbf{\textsc{"Oh my god"}}. The constructed instances and the evaluation results are shown in Figure \ref{fig: ablate}. In general, the designed acoustic and visual mixup and the unimodal tasks guide the model to be aware of the emotion-bearing non-verbal context resulting in more accurate sentiment intensity prediction. Moreover, the modality mixup contributes more to the \#AV-2, \#AV-5, and \#AV-6 instances which might be unseen in the training data. All above results demonstrate that the proposed AV-MC framework can alleviate the text-predominant phenomenon in MSA tasks by perceiving from the non-verbal behaviors. 

\begin{figure}[htbp]
  \centering
  \includegraphics[width=0.8\textwidth]{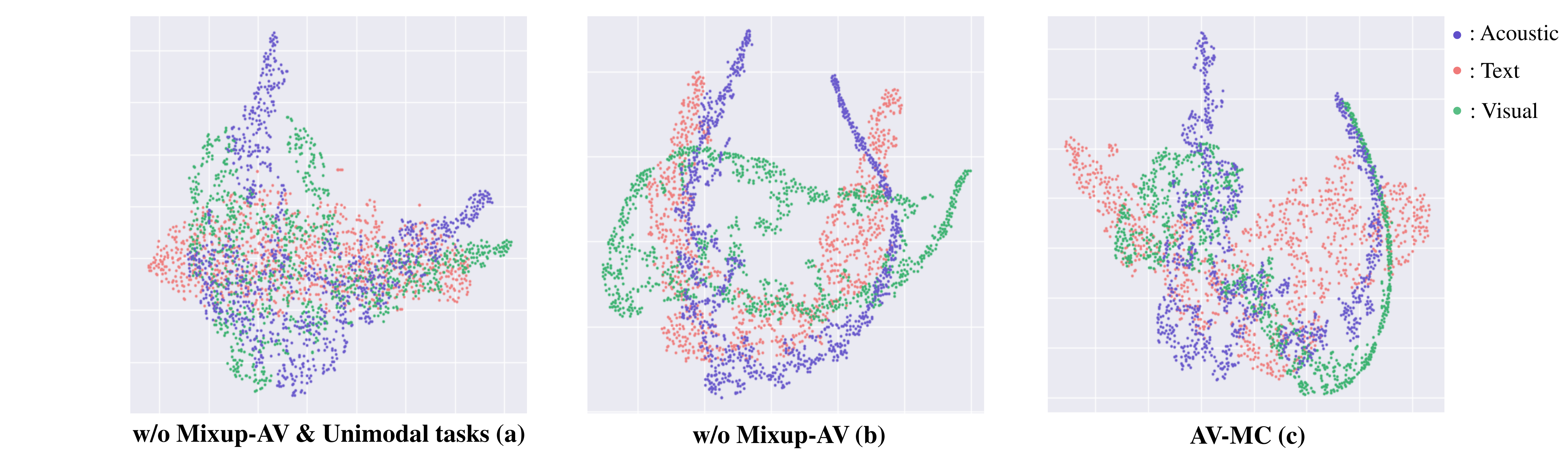}
  \caption{Visualization of unimodal representation distribution for ablation models.}
  \label{fig: tav}
\end{figure}

Furthermore, from a theoretical point of view, the unimodal representation distribution is illustrated with t-SNE \cite{van2008visualizing} to interpret the effect of the modality mixup and unimodal task. The visualization is shown in Figure \ref{fig: tav}. From the comparison between (a) and (b) in Figure, the distribution overlap of different modalities representation is reduced, which indicates that multitask late fusion backbone can preserve the modality-specific information with the help of unimodal annotation. Moreover, from the comparison between (b) and (c) in Figure, the representation distribution within each modality is more separated. Such result indicates that the acoustic and visual mixup strategy can provide more potential non-verbal contexts. The generated virtual contexts are embedded in the latent representation space, enhancing the model's perception ability for the fine-grained sentiment.

\section{Conclusion}
In this work, we strive to improve the contribution of non-verbal cues for sentiment analysis from two aspects. Firstly, from the resource aspect, the CH-SIMS v2.0, a semi-supervised dataset with fine-grained unimodal and multimodal annotations, is constructed by videos with expressive non-verbal behaviors. The noticeable discrepancy between the textual annotation and the multimodal annotation intuitively reveals the importance of making use of emotion-bearing acoustic and visual cues for the prediction. Secondly, from the model aspect, the Acoustic Visual Mixup Consistent Module (AV-MC) is designed. By producing augmentation with mixed acoustic and visual modalities combinations, the non-verbal behaviors are enriched and further help the model be aware of the contributions of acoustic and visual behaviors. The proposed AV-MC module can be regarded as an early attempt based on the CH-SIMS v2.0 dataset. We believe this dataset will guide the research toward real multimodal fusion approaches to alleviate the text-predominant phenomenon. Building upon the CH-SIMS v2.0 dataset, future research can explore the design of interpretable approaches discovering the decisive modalities with unimodal annotations as well as the emotion-bearing multimodal pretrained approaches based on the unsupervised data for multimodal sentiment analysis.
\begin{acks}
% \end{acks}
This paper is funded by The National Natural Science Foundation of China (Grant No. 62173195), Beijing Academy of Artificial Intelligence (BAAI), and Natural Science Foundation of Hebei Province (Grant No. F2022208006). The authors thank the reviewers for their valuable suggestions.
% \begin{acks}
\end{acks}
\newpage

\bibliographystyle{ACM-Reference-Format}

\begin{thebibliography}{48}

%%% the Web address.

\ifx \showCODEN    \undefined \def \showCODEN     #1{\unskip}     \fi
\ifx \showDOI      \undefined \def \showDOI       #1{#1}\fi
\ifx \showISBNx    \undefined \def \showISBNx     #1{\unskip}     \fi
\ifx \showISBNxiii \undefined \def \showISBNxiii  #1{\unskip}     \fi
\ifx \showISSN     \undefined \def \showISSN      #1{\unskip}     \fi
\ifx \showLCCN     \undefined \def \showLCCN      #1{\unskip}     \fi
\ifx \shownote     \undefined \def \shownote      #1{#1}          \fi
\ifx \showarticletitle \undefined \def \showarticletitle #1{#1}   \fi
\ifx \showURL      \undefined \def \showURL       {\relax}        \fi
% invisible to TeX
\providecommand\bibfield[2]{#2}
\providecommand\bibinfo[2]{#2}
\providecommand\natexlab[1]{#1}
\providecommand\showeprint[2][]{arXiv:#2}

\bibitem[Amiriparian et~al\mbox{.}(2021)]%
        {amiriparian2021impact}
\bibfield{author}{\bibinfo{person}{Shahin Amiriparian}, \bibinfo{person}{Artem
  Sokolov}, \bibinfo{person}{Ilhan Aslan}, \bibinfo{person}{Lukas Christ},
  \bibinfo{person}{Maurice Gerczuk}, \bibinfo{person}{Tobias H{\"u}bner},
  \bibinfo{person}{Dmitry Lamanov}, \bibinfo{person}{Manuel Milling},
  \bibinfo{person}{Sandra Ottl}, \bibinfo{person}{Ilya Poduremennykh},
  {et~al\mbox{.}}} \bibinfo{year}{2021}\natexlab{}.
\newblock \showarticletitle{On the Impact of Word Error Rate on
  Acoustic-Linguistic Speech Emotion Recognition: An Update for the Deep
  Learning Era}.
\newblock \bibinfo{journal}{\emph{arXiv preprint arXiv:2104.10121}}
  (\bibinfo{year}{2021}).
\newblock


\bibitem[Baltru{\v{s}}aitis et~al\mbox{.}(2018)]%
        {baltruvsaitis2018multimodal}
\bibfield{author}{\bibinfo{person}{Tadas Baltru{\v{s}}aitis},
  \bibinfo{person}{Chaitanya Ahuja}, {and} \bibinfo{person}{Louis-Philippe
  Morency}.} \bibinfo{year}{2018}\natexlab{}.
\newblock \showarticletitle{Multimodal machine learning: A survey and
  taxonomy}.
\newblock \bibinfo{journal}{\emph{IEEE transactions on pattern analysis and
  machine intelligence}} \bibinfo{volume}{41}, \bibinfo{number}{2}
  (\bibinfo{year}{2018}), \bibinfo{pages}{423--443}.
\newblock


\bibitem[Busso et~al\mbox{.}(2008)]%
        {busso2008iemocap}
\bibfield{author}{\bibinfo{person}{Carlos Busso}, \bibinfo{person}{Murtaza
  Bulut}, \bibinfo{person}{Chi-Chun Lee}, \bibinfo{person}{Abe Kazemzadeh},
  \bibinfo{person}{Emily Mower}, \bibinfo{person}{Samuel Kim},
  \bibinfo{person}{Jeannette~N Chang}, \bibinfo{person}{Sungbok Lee}, {and}
  \bibinfo{person}{Shrikanth~S Narayanan}.} \bibinfo{year}{2008}\natexlab{}.
\newblock \showarticletitle{IEMOCAP: Interactive emotional dyadic motion
  capture database}.
\newblock \bibinfo{journal}{\emph{Language resources and evaluation}}
  \bibinfo{volume}{42}, \bibinfo{number}{4} (\bibinfo{year}{2008}),
  \bibinfo{pages}{335--359}.
\newblock


\bibitem[Chen et~al\mbox{.}(2020)]%
        {2020MixText}
\bibfield{author}{\bibinfo{person}{J. Chen}, \bibinfo{person}{Z. Yang}, {and}
  \bibinfo{person}{D. Yang}.} \bibinfo{year}{2020}\natexlab{}.
\newblock \showarticletitle{MixText: Linguistically-Informed Interpolation of
  Hidden Space for Semi-Supervised Text Classification}.
\newblock  (\bibinfo{year}{2020}).
\newblock


\bibitem[Devlin et~al\mbox{.}(2018)]%
        {devlin2018bert}
\bibfield{author}{\bibinfo{person}{Jacob Devlin}, \bibinfo{person}{Ming-Wei
  Chang}, \bibinfo{person}{Kenton Lee}, {and} \bibinfo{person}{Kristina
  Toutanova}.} \bibinfo{year}{2018}\natexlab{}.
\newblock \showarticletitle{Bert: Pre-training of deep bidirectional
  transformers for language understanding}.
\newblock \bibinfo{journal}{\emph{arXiv preprint arXiv:1810.04805}}
  (\bibinfo{year}{2018}).
\newblock


\bibitem[Eyben et~al\mbox{.}(2015)]%
        {egemaps}
\bibfield{author}{\bibinfo{person}{Florian Eyben}, \bibinfo{person}{Klaus~R
  Scherer}, \bibinfo{person}{Bj{\"o}rn~W Schuller}, \bibinfo{person}{Johan
  Sundberg}, \bibinfo{person}{Elisabeth Andr{\'e}}, \bibinfo{person}{Carlos
  Busso}, \bibinfo{person}{Laurence~Y Devillers}, \bibinfo{person}{Julien
  Epps}, \bibinfo{person}{Petri Laukka}, \bibinfo{person}{Shrikanth~S
  Narayanan}, {et~al\mbox{.}}} \bibinfo{year}{2015}\natexlab{}.
\newblock \showarticletitle{The Geneva minimalistic acoustic parameter set
  (GeMAPS) for voice research and affective computing}.
\newblock \bibinfo{journal}{\emph{IEEE transactions on affective computing}}
  \bibinfo{volume}{7}, \bibinfo{number}{2} (\bibinfo{year}{2015}),
  \bibinfo{pages}{190--202}.
\newblock


\bibitem[Eyben et~al\mbox{.}(2010)]%
        {eyben2010opensmile}
\bibfield{author}{\bibinfo{person}{Florian Eyben}, \bibinfo{person}{Martin
  W{\"o}llmer}, {and} \bibinfo{person}{Bj{\"o}rn Schuller}.}
  \bibinfo{year}{2010}\natexlab{}.
\newblock \showarticletitle{Opensmile: the munich versatile and fast
  open-source audio feature extractor}. In
  \bibinfo{booktitle}{\emph{Proceedings of the 18th ACM international
  conference on Multimedia}}. \bibinfo{pages}{1459--1462}.
\newblock


\bibitem[Ghosal et~al\mbox{.}(2020)]%
        {ghosal2020utterance}
\bibfield{author}{\bibinfo{person}{Deepanway Ghosal}, \bibinfo{person}{Navonil
  Majumder}, \bibinfo{person}{Rada Mihalcea}, {and} \bibinfo{person}{Soujanya
  Poria}.} \bibinfo{year}{2020}\natexlab{}.
\newblock \showarticletitle{Utterance-level dialogue understanding: An
  empirical study}.
\newblock \bibinfo{journal}{\emph{arXiv preprint arXiv:2009.13902}}
  (\bibinfo{year}{2020}).
\newblock


\bibitem[Guo et~al\mbox{.}(2019)]%
        {2019Augmenting}
\bibfield{author}{\bibinfo{person}{H. Guo}, \bibinfo{person}{Y. Mao}, {and}
  \bibinfo{person}{R. Zhang}.} \bibinfo{year}{2019}\natexlab{}.
\newblock \showarticletitle{Augmenting Data with Mixup for Sentence
  Classification: An Empirical Study}.
\newblock  (\bibinfo{year}{2019}).
\newblock


\bibitem[Han et~al\mbox{.}(2021)]%
        {MMIM}
\bibfield{author}{\bibinfo{person}{Wei Han}, \bibinfo{person}{Hui Chen}, {and}
  \bibinfo{person}{Soujanya Poria}.} \bibinfo{year}{2021}\natexlab{}.
\newblock \showarticletitle{Improving Multimodal Fusion with Hierarchical
  Mutual Information Maximization for Multimodal Sentiment Analysis}.
\newblock \bibinfo{journal}{\emph{arXiv preprint arXiv:2109.00412}}
  (\bibinfo{year}{2021}).
\newblock


\bibitem[Hazarika et~al\mbox{.}(2020)]%
        {misa}
\bibfield{author}{\bibinfo{person}{Devamanyu Hazarika}, \bibinfo{person}{Roger
  Zimmermann}, {and} \bibinfo{person}{Soujanya Poria}.}
  \bibinfo{year}{2020}\natexlab{}.
\newblock \showarticletitle{{MISA:} Modality-Invariant and -Specific
  Representations for Multimodal Sentiment Analysis}.
\newblock \bibinfo{journal}{\emph{CoRR}}  \bibinfo{volume}{abs/2005.03545}
  (\bibinfo{year}{2020}).
\newblock
\showeprint[arxiv]{2005.03545}
\urldef\tempurl%
\url{https://arxiv.org/abs/2005.03545}
\showURL{%
\tempurl}


\bibitem[Hochreiter and Schmidhuber(1997)]%
        {1997Long}
\bibfield{author}{\bibinfo{person}{S Hochreiter} {and} \bibinfo{person}{J
  Schmidhuber}.} \bibinfo{year}{1997}\natexlab{}.
\newblock \showarticletitle{Long Short-Term Memory}.
\newblock \bibinfo{journal}{\emph{Neural Computation}} \bibinfo{volume}{9},
  \bibinfo{number}{8} (\bibinfo{year}{1997}), \bibinfo{pages}{1735--1780}.
\newblock


\bibitem[Li and Chen(2020)]%
        {li2020multimodal}
\bibfield{author}{\bibinfo{person}{Xia Li} {and} \bibinfo{person}{Minping
  Chen}.} \bibinfo{year}{2020}\natexlab{}.
\newblock \showarticletitle{Multimodal sentiment analysis with
  multi-perspective fusion network focusing on sense attentive language}. In
  \bibinfo{booktitle}{\emph{China National Conference on Chinese Computational
  Linguistics}}. Springer, \bibinfo{pages}{359--373}.
\newblock


\bibitem[Li et~al\mbox{.}(2017)]%
        {li2017cheavd}
\bibfield{author}{\bibinfo{person}{Ya Li}, \bibinfo{person}{Jianhua Tao},
  \bibinfo{person}{Linlin Chao}, \bibinfo{person}{Wei Bao}, {and}
  \bibinfo{person}{Yazhu Liu}.} \bibinfo{year}{2017}\natexlab{}.
\newblock \showarticletitle{CHEAVD: a Chinese natural emotional audio--visual
  database}.
\newblock \bibinfo{journal}{\emph{Journal of Ambient Intelligence and Humanized
  Computing}} \bibinfo{volume}{8}, \bibinfo{number}{6} (\bibinfo{year}{2017}),
  \bibinfo{pages}{913--924}.
\newblock


\bibitem[Liang et~al\mbox{.}(2021)]%
        {liang2021multibench}
\bibfield{author}{\bibinfo{person}{Paul~Pu Liang}, \bibinfo{person}{Yiwei Lyu},
  \bibinfo{person}{Xiang Fan}, \bibinfo{person}{Zetian Wu},
  \bibinfo{person}{Yun Cheng}, \bibinfo{person}{Jason Wu},
  \bibinfo{person}{Leslie Chen}, \bibinfo{person}{Peter Wu},
  \bibinfo{person}{Michelle~A Lee}, \bibinfo{person}{Yuke Zhu},
  {et~al\mbox{.}}} \bibinfo{year}{2021}\natexlab{}.
\newblock \showarticletitle{Multibench: Multiscale benchmarks for multimodal
  representation learning}.
\newblock \bibinfo{journal}{\emph{arXiv preprint arXiv:2107.07502}}
  (\bibinfo{year}{2021}).
\newblock


\bibitem[Liesting et~al\mbox{.}(2021)]%
        {liesting2021data}
\bibfield{author}{\bibinfo{person}{Tomas Liesting}, \bibinfo{person}{Flavius
  Frasincar}, {and} \bibinfo{person}{Maria~Mihaela Tru{\c{s}}c{\u{a}}}.}
  \bibinfo{year}{2021}\natexlab{}.
\newblock \showarticletitle{Data augmentation in a hybrid approach for
  aspect-based sentiment analysis}. In \bibinfo{booktitle}{\emph{Proceedings of
  the 36th Annual ACM Symposium on Applied Computing}}.
  \bibinfo{pages}{828--835}.
\newblock


\bibitem[Liu et~al\mbox{.}(2018)]%
        {lmf}
\bibfield{author}{\bibinfo{person}{Zhun Liu}, \bibinfo{person}{Ying Shen},
  \bibinfo{person}{Varun~Bharadhwaj Lakshminarasimhan},
  \bibinfo{person}{Paul~Pu Liang}, \bibinfo{person}{Amir Zadeh}, {and}
  \bibinfo{person}{Louis-Philippe Morency}.} \bibinfo{year}{2018}\natexlab{}.
\newblock \showarticletitle{Efficient low-rank multimodal fusion with
  modality-specific factors}.
\newblock \bibinfo{journal}{\emph{arXiv preprint arXiv:1806.00064}}
  (\bibinfo{year}{2018}).
\newblock


\bibitem[Luo et~al\mbox{.}(2021)]%
        {luo2021scalevlad}
\bibfield{author}{\bibinfo{person}{Huaishao Luo}, \bibinfo{person}{Lei Ji},
  \bibinfo{person}{Yanyong Huang}, \bibinfo{person}{Bin Wang},
  \bibinfo{person}{Shenggong Ji}, {and} \bibinfo{person}{Tianrui Li}.}
  \bibinfo{year}{2021}\natexlab{}.
\newblock \showarticletitle{ScaleVLAD: Improving Multimodal Sentiment Analysis
  via Multi-Scale Fusion of Locally Descriptors}.
\newblock \bibinfo{journal}{\emph{arXiv preprint arXiv:2112.01368}}
  (\bibinfo{year}{2021}).
\newblock


\bibitem[Ma et~al\mbox{.}(2020)]%
        {ma2020survey}
\bibfield{author}{\bibinfo{person}{Yukun Ma}, \bibinfo{person}{Khanh~Linh
  Nguyen}, \bibinfo{person}{Frank~Z Xing}, {and} \bibinfo{person}{Erik
  Cambria}.} \bibinfo{year}{2020}\natexlab{}.
\newblock \showarticletitle{A survey on empathetic dialogue systems}.
\newblock \bibinfo{journal}{\emph{Information Fusion}}  \bibinfo{volume}{64}
  (\bibinfo{year}{2020}), \bibinfo{pages}{50--70}.
\newblock


\bibitem[Mao et~al\mbox{.}(2022)]%
        {mao2022m}
\bibfield{author}{\bibinfo{person}{Huisheng Mao}, \bibinfo{person}{Ziqi Yuan},
  \bibinfo{person}{Hua Xu}, \bibinfo{person}{Wenmeng Yu}, \bibinfo{person}{Yihe
  Liu}, {and} \bibinfo{person}{Kai Gao}.} \bibinfo{year}{2022}\natexlab{}.
\newblock \showarticletitle{M-SENA: An Integrated Platform for Multimodal
  Sentiment Analysis}. In \bibinfo{booktitle}{\emph{Proceedings of the 60th
  Annual Meeting of the Association for Computational Linguistics: System
  Demonstrations}}. \bibinfo{pages}{204--213}.
\newblock


\bibitem[Morency et~al\mbox{.}(2011)]%
        {morency2011towards}
\bibfield{author}{\bibinfo{person}{Louis-Philippe Morency},
  \bibinfo{person}{Rada Mihalcea}, {and} \bibinfo{person}{Payal Doshi}.}
  \bibinfo{year}{2011}\natexlab{}.
\newblock \showarticletitle{Towards multimodal sentiment analysis: Harvesting
  opinions from the web}. In \bibinfo{booktitle}{\emph{Proceedings of the 13th
  international conference on multimodal interfaces}}.
  \bibinfo{pages}{169--176}.
\newblock


\bibitem[Peng et~al\mbox{.}(2022)]%
        {peng2022balanced}
\bibfield{author}{\bibinfo{person}{Xiaokang Peng}, \bibinfo{person}{Yake Wei},
  \bibinfo{person}{Andong Deng}, \bibinfo{person}{Dong Wang}, {and}
  \bibinfo{person}{Di Hu}.} \bibinfo{year}{2022}\natexlab{}.
\newblock \showarticletitle{Balanced Multimodal Learning via On-the-fly
  Gradient Modulation}.
\newblock \bibinfo{journal}{\emph{arXiv preprint arXiv:2203.15332}}
  (\bibinfo{year}{2022}).
\newblock


\bibitem[Poria et~al\mbox{.}(2020)]%
        {poria2020beneath}
\bibfield{author}{\bibinfo{person}{Soujanya Poria}, \bibinfo{person}{Devamanyu
  Hazarika}, \bibinfo{person}{Navonil Majumder}, {and} \bibinfo{person}{Rada
  Mihalcea}.} \bibinfo{year}{2020}\natexlab{}.
\newblock \showarticletitle{Beneath the tip of the iceberg: Current challenges
  and new directions in sentiment analysis research}.
\newblock \bibinfo{journal}{\emph{IEEE Transactions on Affective Computing}}
  (\bibinfo{year}{2020}).
\newblock


\bibitem[Poria et~al\mbox{.}(2019)]%
        {2019MELD}
\bibfield{author}{\bibinfo{person}{S. Poria}, \bibinfo{person}{D. Hazarika},
  \bibinfo{person}{N. Majumder}, \bibinfo{person}{G. Naik}, {and}
  \bibinfo{person}{R. Mihalcea}.} \bibinfo{year}{2019}\natexlab{}.
\newblock \showarticletitle{MELD: A Multimodal Multi-Party Dataset for Emotion
  Recognition in Conversations}. In \bibinfo{booktitle}{\emph{Proceedings of
  the 57th Annual Meeting of the Association for Computational Linguistics}}.
\newblock


\bibitem[Poria et~al\mbox{.}(2018)]%
        {poria2018multimodal}
\bibfield{author}{\bibinfo{person}{Soujanya Poria}, \bibinfo{person}{Navonil
  Majumder}, \bibinfo{person}{Devamanyu Hazarika}, \bibinfo{person}{Erik
  Cambria}, \bibinfo{person}{Alexander Gelbukh}, {and} \bibinfo{person}{Amir
  Hussain}.} \bibinfo{year}{2018}\natexlab{}.
\newblock \showarticletitle{Multimodal sentiment analysis: Addressing key
  issues and setting up the baselines}.
\newblock \bibinfo{journal}{\emph{IEEE Intelligent Systems}}
  \bibinfo{volume}{33}, \bibinfo{number}{6} (\bibinfo{year}{2018}),
  \bibinfo{pages}{17--25}.
\newblock


\bibitem[Rahman et~al\mbox{.}(2020)]%
        {BERT_MAG}
\bibfield{author}{\bibinfo{person}{Wasifur Rahman}, \bibinfo{person}{Md~Kamrul
  Hasan}, \bibinfo{person}{Sangwu Lee}, \bibinfo{person}{AmirAli~Bagher Zadeh},
  \bibinfo{person}{Chengfeng Mao}, \bibinfo{person}{Louis-Philippe Morency},
  {and} \bibinfo{person}{Ehsan Hoque}.} \bibinfo{year}{2020}\natexlab{}.
\newblock \showarticletitle{Integrating multimodal information in large
  pretrained transformers}. In \bibinfo{booktitle}{\emph{Proceedings of the
  58th Annual Meeting of the Association for Computational Linguistics}}.
  \bibinfo{pages}{2359--2369}.
\newblock


\bibitem[Tao and Tan(2005)]%
        {tao2005affective}
\bibfield{author}{\bibinfo{person}{Jianhua Tao} {and} \bibinfo{person}{Tieniu
  Tan}.} \bibinfo{year}{2005}\natexlab{}.
\newblock \showarticletitle{Affective computing: A review}. In
  \bibinfo{booktitle}{\emph{International Conference on Affective computing and
  intelligent interaction}}. Springer, \bibinfo{pages}{981--995}.
\newblock


\bibitem[Tao et~al\mbox{.}(2021)]%
        {tao2021someone}
\bibfield{author}{\bibinfo{person}{Ruijie Tao}, \bibinfo{person}{Zexu Pan},
  \bibinfo{person}{Rohan~Kumar Das}, \bibinfo{person}{Xinyuan Qian},
  \bibinfo{person}{Mike~Zheng Shou}, {and} \bibinfo{person}{Haizhou Li}.}
  \bibinfo{year}{2021}\natexlab{}.
\newblock \showarticletitle{Is Someone Speaking? Exploring Long-term Temporal
  Features for Audio-visual Active Speaker Detection}. In
  \bibinfo{booktitle}{\emph{Proceedings of the 29th ACM International
  Conference on Multimedia}}. \bibinfo{pages}{3927–3935}.
\newblock


\bibitem[{Tsai} et~al\mbox{.}(2019)]%
        {mult}
\bibfield{author}{\bibinfo{person}{Yao-Hung~Hubert {Tsai}},
  \bibinfo{person}{Shaojie {Bai}}, \bibinfo{person}{Paul~Pu {Liang}},
  \bibinfo{person}{J.~Zico {Kolter}}, \bibinfo{person}{Louis-Philippe
  {Morency}}, {and} \bibinfo{person}{Ruslan {Salakhutdinov}}.}
  \bibinfo{year}{2019}\natexlab{}.
\newblock \showarticletitle{Multimodal Transformer for Unaligned Multimodal
  Language Sequences}. In \bibinfo{booktitle}{\emph{Proceedings of the 57th
  Annual Meeting of the Association for Computational Linguistics}},
  Vol.~\bibinfo{volume}{2019}. \bibinfo{pages}{6558--6569}.
\newblock


\bibitem[Uddin et~al\mbox{.}(2020)]%
        {2020SaliencyMix}
\bibfield{author}{\bibinfo{person}{Afms Uddin}, \bibinfo{person}{M.~S. Monira},
  \bibinfo{person}{W. Shin}, \bibinfo{person}{T.~C. Chung}, {and}
  \bibinfo{person}{S.~H. Bae}.} \bibinfo{year}{2020}\natexlab{}.
\newblock \showarticletitle{SaliencyMix: A Saliency Guided Data Augmentation
  Strategy for Better Regularization}.
\newblock  (\bibinfo{year}{2020}).
\newblock


\bibitem[Van~der Maaten and Hinton(2008)]%
        {van2008visualizing}
\bibfield{author}{\bibinfo{person}{Laurens Van~der Maaten} {and}
  \bibinfo{person}{Geoffrey Hinton}.} \bibinfo{year}{2008}\natexlab{}.
\newblock \showarticletitle{Visualizing data using t-SNE.}
\newblock \bibinfo{journal}{\emph{Journal of machine learning research}}
  \bibinfo{volume}{9}, \bibinfo{number}{11} (\bibinfo{year}{2008}).
\newblock


\bibitem[Verma et~al\mbox{.}(2019)]%
        {verma2019interpolation}
\bibfield{author}{\bibinfo{person}{Vikas Verma}, \bibinfo{person}{Kenji
  Kawaguchi}, \bibinfo{person}{Alex Lamb}, \bibinfo{person}{Juho Kannala},
  \bibinfo{person}{Yoshua Bengio}, {and} \bibinfo{person}{David Lopez-Paz}.}
  \bibinfo{year}{2019}\natexlab{}.
\newblock \showarticletitle{Interpolation consistency training for
  semi-supervised learning}.
\newblock \bibinfo{journal}{\emph{arXiv preprint arXiv:1903.03825}}
  (\bibinfo{year}{2019}).
\newblock


\bibitem[Verma et~al\mbox{.}(2018)]%
        {2018Manifold}
\bibfield{author}{\bibinfo{person}{V. Verma}, \bibinfo{person}{A. Lamb},
  \bibinfo{person}{C. Beckham}, \bibinfo{person}{A. Najafi},
  \bibinfo{person}{I. Mitliagkas}, \bibinfo{person}{A. Courville},
  \bibinfo{person}{D. Lopez-Paz}, {and} \bibinfo{person}{Y. Bengio}.}
  \bibinfo{year}{2018}\natexlab{}.
\newblock \showarticletitle{Manifold Mixup: Better Representations by
  Interpolating Hidden States}.
\newblock  (\bibinfo{year}{2018}).
\newblock


\bibitem[Vva et~al\mbox{.}(2021)]%
        {2021Interpolation}
\bibfield{author}{\bibinfo{person}{B Vva}, \bibinfo{person}{C Kk},
  \bibinfo{person}{A Al}, \bibinfo{person}{B Jk}, \bibinfo{person}{B As},
  \bibinfo{person}{A Yb}, {and} \bibinfo{person}{D Lp}.}
  \bibinfo{year}{2021}\natexlab{}.
\newblock \showarticletitle{Interpolation consistency training for
  semi-supervised learning}.
\newblock  (\bibinfo{year}{2021}).
\newblock


\bibitem[Wang et~al\mbox{.}(2019)]%
        {wang2019words}
\bibfield{author}{\bibinfo{person}{Yansen Wang}, \bibinfo{person}{Ying Shen},
  \bibinfo{person}{Zhun Liu}, \bibinfo{person}{Paul~Pu Liang},
  \bibinfo{person}{Amir Zadeh}, {and} \bibinfo{person}{Louis-Philippe
  Morency}.} \bibinfo{year}{2019}\natexlab{}.
\newblock \showarticletitle{Words can shift: Dynamically adjusting word
  representations using nonverbal behaviors}. In
  \bibinfo{booktitle}{\emph{Proceedings of the AAAI Conference on Artificial
  Intelligence}}, Vol.~\bibinfo{volume}{33}. \bibinfo{pages}{7216--7223}.
\newblock


\bibitem[Williams et~al\mbox{.}(2018)]%
        {lf_dnn_ef_lstm}
\bibfield{author}{\bibinfo{person}{Jennifer Williams}, \bibinfo{person}{Steven
  Kleinegesse}, \bibinfo{person}{Ramona Comanescu}, {and} \bibinfo{person}{Oana
  Radu}.} \bibinfo{year}{2018}\natexlab{}.
\newblock \showarticletitle{Recognizing Emotions in Video Using Multimodal DNN
  Feature Fusion}. In \bibinfo{booktitle}{\emph{Proceedings of Grand Challenge
  and Workshop on Human Multimodal Language (Challenge-HML)}}.
  \bibinfo{publisher}{Association for Computational Linguistics},
  \bibinfo{pages}{11--19}.
\newblock


\bibitem[Wu et~al\mbox{.}(2022)]%
        {wu2022sentiment}
\bibfield{author}{\bibinfo{person}{Yang Wu}, \bibinfo{person}{Yanyan Zhao},
  \bibinfo{person}{Hao Yang}, \bibinfo{person}{Song Chen},
  \bibinfo{person}{Bing Qin}, \bibinfo{person}{Xiaohuan Cao}, {and}
  \bibinfo{person}{Wenting Zhao}.} \bibinfo{year}{2022}\natexlab{}.
\newblock \showarticletitle{Sentiment Word Aware Multimodal Refinement for
  Multimodal Sentiment Analysis with ASR Errors}.
\newblock \bibinfo{journal}{\emph{arXiv preprint arXiv:2203.00257}}
  (\bibinfo{year}{2022}).
\newblock


\bibitem[Yoon et~al\mbox{.}(2021)]%
        {2021SSMix}
\bibfield{author}{\bibinfo{person}{S. Yoon}, \bibinfo{person}{G. Kim}, {and}
  \bibinfo{person}{K. Park}.} \bibinfo{year}{2021}\natexlab{}.
\newblock \showarticletitle{SSMix: Saliency-Based Span Mixup for Text
  Classification}.
\newblock  (\bibinfo{year}{2021}).
\newblock


\bibitem[Yu et~al\mbox{.}(2020)]%
        {yu2020ch}
\bibfield{author}{\bibinfo{person}{Wenmeng Yu}, \bibinfo{person}{Hua Xu},
  \bibinfo{person}{Fanyang Meng}, \bibinfo{person}{Yilin Zhu},
  \bibinfo{person}{Yixiao Ma}, \bibinfo{person}{Jiele Wu},
  \bibinfo{person}{Jiyun Zou}, {and} \bibinfo{person}{Kaicheng Yang}.}
  \bibinfo{year}{2020}\natexlab{}.
\newblock \showarticletitle{Ch-sims: A chinese multimodal sentiment analysis
  dataset with fine-grained annotation of modality}. In
  \bibinfo{booktitle}{\emph{Proceedings of the 58th Annual Meeting of the
  Association for Computational Linguistics}}. \bibinfo{pages}{3718--3727}.
\newblock


\bibitem[Yu et~al\mbox{.}(2021)]%
        {self-mm}
\bibfield{author}{\bibinfo{person}{Wenmeng Yu}, \bibinfo{person}{Hua Xu},
  \bibinfo{person}{Ziqi Yuan}, {and} \bibinfo{person}{Jiele Wu}.}
  \bibinfo{year}{2021}\natexlab{}.
\newblock \showarticletitle{Learning Modality-Specific Representations with
  Self-Supervised Multi-Task Learning for Multimodal Sentiment Analysis}.
\newblock \bibinfo{journal}{\emph{arXiv preprint arXiv:2102.04830}}
  (\bibinfo{year}{2021}).
\newblock


\bibitem[Yun et~al\mbox{.}([n.\,d.])]%
        {0CutMix}
\bibfield{author}{\bibinfo{person}{S. Yun}, \bibinfo{person}{D. Han},
  \bibinfo{person}{S. Chun}, \bibinfo{person}{S.~J. Oh}, \bibinfo{person}{Y.
  Yoo}, {and} \bibinfo{person}{J. Choe}.} \bibinfo{year}{[n.\,d.]}\natexlab{}.
\newblock \showarticletitle{CutMix: Regularization Strategy to Train Strong
  Classifiers With Localizable Features}. In
  \bibinfo{booktitle}{\emph{International Conference on Computer Vision}}.
\newblock


\bibitem[Zadeh et~al\mbox{.}(2020)]%
        {zadeh2020cmu}
\bibfield{author}{\bibinfo{person}{Amir Zadeh}, \bibinfo{person}{Yan~Sheng
  Cao}, \bibinfo{person}{Simon Hessner}, \bibinfo{person}{Paul~Pu Liang},
  \bibinfo{person}{Soujanya Poria}, {and} \bibinfo{person}{Louis-Philippe
  Morency}.} \bibinfo{year}{2020}\natexlab{}.
\newblock \showarticletitle{CMU-MOSEAS: A multimodal language dataset for
  Spanish, Portuguese, German and French}. In
  \bibinfo{booktitle}{\emph{Proceedings of the Conference on Empirical Methods
  in Natural Language Processing. Conference on Empirical Methods in Natural
  Language Processing}}, Vol.~\bibinfo{volume}{2020}. NIH Public Access,
  \bibinfo{pages}{1801}.
\newblock


\bibitem[Zadeh et~al\mbox{.}(2017)]%
        {tfn}
\bibfield{author}{\bibinfo{person}{Amir Zadeh}, \bibinfo{person}{Minghai Chen},
  \bibinfo{person}{Soujanya Poria}, \bibinfo{person}{Erik Cambria}, {and}
  \bibinfo{person}{Louis-Philippe Morency}.} \bibinfo{year}{2017}\natexlab{}.
\newblock \showarticletitle{Tensor fusion network for multimodal sentiment
  analysis}.
\newblock \bibinfo{journal}{\emph{arXiv preprint arXiv:1707.07250}}
  (\bibinfo{year}{2017}).
\newblock


\bibitem[Zadeh et~al\mbox{.}(2018a)]%
        {mfn}
\bibfield{author}{\bibinfo{person}{Amir Zadeh}, \bibinfo{person}{Paul~Pu
  Liang}, \bibinfo{person}{Navonil Mazumder}, \bibinfo{person}{Soujanya Poria},
  \bibinfo{person}{Erik Cambria}, {and} \bibinfo{person}{Louis-Philippe
  Morency}.} \bibinfo{year}{2018}\natexlab{a}.
\newblock \showarticletitle{Memory fusion network for multi-view sequential
  learning}.
\newblock \bibinfo{journal}{\emph{arXiv preprint arXiv:1802.00927}}
  (\bibinfo{year}{2018}).
\newblock


\bibitem[Zadeh et~al\mbox{.}(2016)]%
        {mosi}
\bibfield{author}{\bibinfo{person}{Amir Zadeh}, \bibinfo{person}{Rowan
  Zellers}, \bibinfo{person}{Eli Pincus}, {and} \bibinfo{person}{Louis-Philippe
  Morency}.} \bibinfo{year}{2016}\natexlab{}.
\newblock \showarticletitle{Multimodal sentiment intensity analysis in videos:
  Facial gestures and verbal messages}.
\newblock \bibinfo{journal}{\emph{IEEE Intelligent Systems}}
  \bibinfo{volume}{31}, \bibinfo{number}{6} (\bibinfo{year}{2016}),
  \bibinfo{pages}{82--88}.
\newblock


\bibitem[Zadeh et~al\mbox{.}(2018b)]%
        {mosei}
\bibfield{author}{\bibinfo{person}{AmirAli~Bagher Zadeh},
  \bibinfo{person}{Paul~Pu Liang}, \bibinfo{person}{Soujanya Poria},
  \bibinfo{person}{Erik Cambria}, {and} \bibinfo{person}{Louis-Philippe
  Morency}.} \bibinfo{year}{2018}\natexlab{b}.
\newblock \showarticletitle{Multimodal language analysis in the wild: Cmu-mosei
  dataset and interpretable dynamic fusion graph}. In
  \bibinfo{booktitle}{\emph{Proceedings of the 56th Annual Meeting of the
  Association for Computational Linguistics (Volume 1: Long Papers)}}.
  \bibinfo{pages}{2236--2246}.
\newblock


\bibitem[Zhang et~al\mbox{.}(2017)]%
        {zhang2017mixup}
\bibfield{author}{\bibinfo{person}{H Zhang}, \bibinfo{person}{M Cisse},
  \bibinfo{person}{YN Dauphin}, {and} \bibinfo{person}{D Lopez-Paz}.}
  \bibinfo{year}{2017}\natexlab{}.
\newblock \bibinfo{title}{Mixup: beyond empirical risk minimization.
  International Conference on Learning Representations}.
\newblock
\newblock


\bibitem[Zhang et~al\mbox{.}(2016)]%
        {zhang2016joint}
\bibfield{author}{\bibinfo{person}{Kaipeng Zhang}, \bibinfo{person}{Zhanpeng
  Zhang}, \bibinfo{person}{Zhifeng Li}, {and} \bibinfo{person}{Yu Qiao}.}
  \bibinfo{year}{2016}\natexlab{}.
\newblock \showarticletitle{Joint face detection and alignment using multitask
  cascaded convolutional networks}.
\newblock \bibinfo{journal}{\emph{IEEE signal processing letters}}
  \bibinfo{volume}{23}, \bibinfo{number}{10} (\bibinfo{year}{2016}),
  \bibinfo{pages}{1499--1503}.
\newblock


\end{thebibliography}

\clearpage
\appendix

\end{document}